\documentclass[a4paper,fleqn,usenatbib]{mnras}

\usepackage{mathptmx}

\usepackage[T1]{fontenc}
\usepackage{ae,aecompl}

\usepackage{graphicx}	
\usepackage{amsmath}	
\usepackage{amssymb}	

\title[Calibrations of SITELLE's first data release]{Calibrations of
  SITELLE's first data release}

\author[T. Martin et al.]{
T. Martin$^{1,2}$\thanks{E-mail: thomas.martin.1@ulaval.ca},
L. Drissen$^{1,2}$
\\
$^{1}$D\'epartement de physique, de g\'enie physique et d'optique,
Universit{\'e} Laval, 2325, rue de l'universit{\'e}, Qu{\'e}bec
(Qu{\'e}bec), G1V 0A6, Canada\\
$^{2}$Centre de Recherche en Astrophysique du Qu{\'e}bec
(CRAQ), Montr{\'e}al (Qu{\'e}bec), H3C 3J7, Canada}

\date{Accepted XXX. Received YYY; in original form ZZZ}

\pubyear{2017}

\DeclareRobustCommand{\HII}{\textup{H\,\textsc{\lowercase{II}}}}

\DeclareRobustCommand{\kms}{km\,s$^{-1}$}
\begin{document}
\label{firstpage}
\pagerange{\pageref{firstpage}--\pageref{lastpage}}
\maketitle

\begin{abstract}
  SITELLE is an imaging Fourier Transform Spectrometer installed at
  the Canada-France-Hawaii Telescope since July 2015. It delivers
  spectral cubes covering an 11\arcmin$\times$11\arcmin{} field-of-view
  with a seeing-limited spatial resolution and a tunable spectral
  resolution (R$\sim$1--10\,000) in selected passbands of the
  visible band (350--900\,nm). We present an accurate picture of the
  calibration accuracy of SITELLE's first data release. To this
  purpose, most of the operations of the reduction pipeline (ORBS) are
  described in details.
\end{abstract}




\section{Introduction}

\begin{figure}
  \includegraphics[width=\columnwidth]{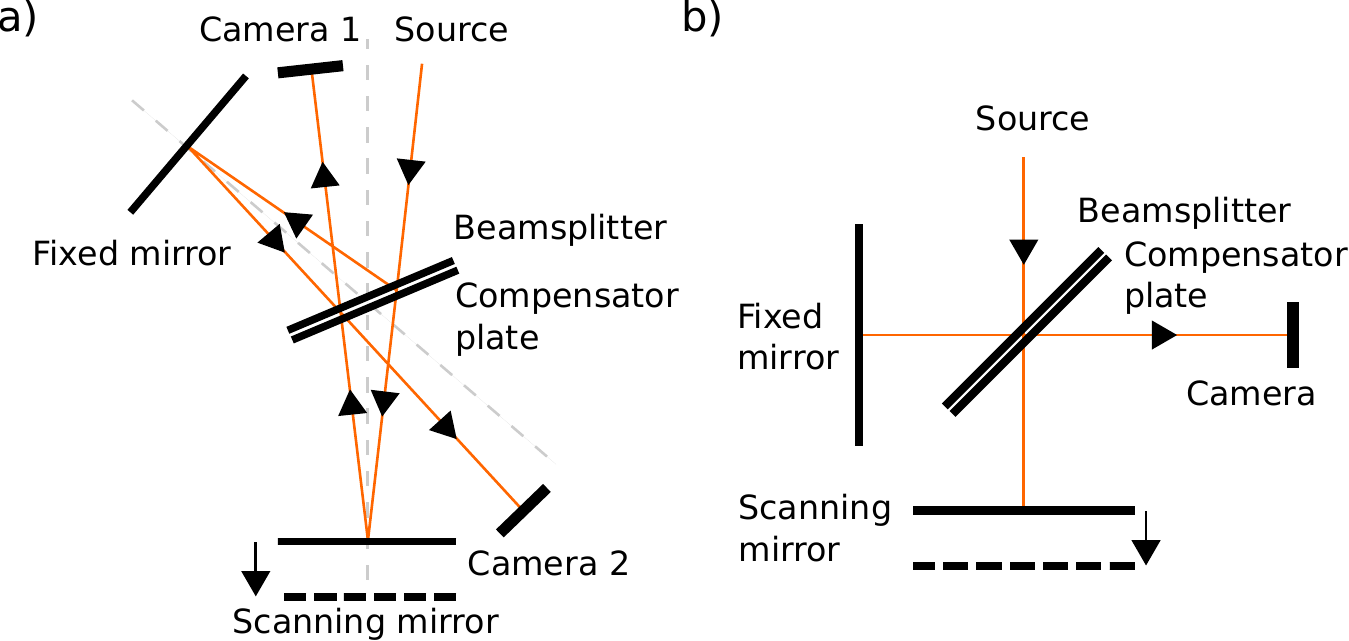}
  \caption{a) Sketch of SITELLE's interferometer. b) Sketch of a
    classical Michelson interferometer.}
    \label{fig:michelson}
\end{figure}

SITELLE \citep{Drissen2010} is an imaging Fourier Transform
Spectrometer installed at the Canada-France-Hawaii Telescope (CFHT)
since July 2015. It delivers spectral cubes covering an
11\arcmin$\times$11\arcmin{} field-of-view with a seeing-limited
spatial resolution and a tunable spectral resolution
(R$\sim$1--10\,000) in selected passbands of the visible band
(350--900\,nm).  SITELLE is based on an off-axis configuration of the
interferometer which gives the opportunity to measure the flux on both
output ports instead of one in the classical, on-axis, configuration
(see Fig.~\ref{fig:michelson}, \citealt{Grandmont2012}). Two
2k$\times$2k CCD cameras are attached on the output ports. During an
observing scan, the scanning mirror is moved. The optical path
difference (OPD) between the two interfering beams is thus gradually
changed, step by step, modulating the output intensity according to
the spectral energy distribution of the observed source. The images
collected at the output are interferometric images which are stacked
into two interferometric cubes (one for each output port). For a given
pixel in the image, the intensity recorded at each step is thus an
interferogram, the Fourier transform of which must be computed to
determine the spectrum of the source. More details on the observing
process are given in \citet{Martin2016}, \citet{Drissen2010} and references
therein.

These two inteferometric cubes are transformed into one spectral cube
with the reduction pipeline ORBS. The details of the operations of the
pipeline which have an impact on the calibrations will be discussed in
this article. The general description of the pipeline is given in
\citet{Martin2012, Martin2015} and \citet{Martin2015t}\footnote{ORBS
  source code can be found at
  \url{https://sourceforge.net/projects/orb-orbs/}.}.

Ideally, the Fourier transform of all the interferograms recorded
(around 4 million) would be the only operation required to transform
an interferometric cube into a spectral cube of the same dimension. In
practice however, a lot of operations on the raw data are actually
needed to prepare and merge the interferograms obtained with each
camera before being able to compute their spectrum and calibrate the
obtained spectral cube. The main focus of this article is to give an
accurate picture of the calibration accuracy of SITELLE's first data
release. Note that this calibration will certainly be improved with
the next data releases. Some of them are described in Martin et
al. (in preparation). Section~\ref{sec:reduction_process} gives an
overview of the reduction pipeline. In
section~\ref{sec:phase_correction} we describe in details the phase
correction process which can have a major impact on the flux and the
wavelength calibrations. Flux calibration, wavelength calibration and
astrometric calibration are considered respectively in
sections~\ref{sec:flux_calibration}, ~\ref{sec:wavelength_calibration}
and~\ref{sec:astrometric_calibration}.

Note that some of the most important results of this paper have
already been presented in a conference proceedings \citep{martin2016b}
but in a much less detailed fashion.

\section{Reduction process}
\label{sec:reduction_process}

Data calibration is the last step of the reduction pipeline. But each
step of the reduction has an impact on the quality of the calibration,
and especially the flux calibration. There has been some changes since
the description of the pipeline given in \citet{Martin2012,
  Martin2015, Martin2015t}. We present here the up-to-date version of
the general processing pipeline:
\begin{enumerate}
\item Calibration of the CCD images: bias is computed from the
  overscan columns and subtracted, flat-field illumination errors are
  corrected. No dark is subtracted because the dark current is very
  low ($<2$\,e$^{-}$/hr\footnote{See the specification of the e2v's
    back-illuminated 2k$\times$2k scientific CCD sensor CCD230-42 at
    \url{http://www.e2v.com/resources/account/download-datasheet/3828}})
  and in this case dark subtraction would enhance the noise in the
  images without having a significant impact on the quality of the
  flux calibration (see section~\ref{sec:image_correction}). Images
  are also realigned to compensate for guiding errors
\item Alignment of the cubes. There is a slight optical misalignment
  between the cameras. As this alignment is expected to change when
  the cameras are removed, the stars detected in the field-of-view are
  used to compute the alignment parameters for each cube.
\item Combination of the cubes. The two inteferometric cubes are
  combined to enhance the SNR, correct for transmission variations of
  the sky during the observation and remove the unmodulated background
  light (see section~\ref{sec:cube_combination}).
\item Phase computation of the combined cube (see
  section~\ref{sec:phase_correction}).
\item Transformation of the combined cube to a spectral cube. The
  transformation is based on a discrete Fourier transform (DFT) of
  each inteferogram. The phase information obtained from the phase
  cube is used at this point.
\item Calibration of the spectral cube (see
  sections~\ref{sec:flux_calibration},
  ~\ref{sec:wavelength_calibration}
  and~\ref{sec:astrometric_calibration}).
\end{enumerate}
The impact of these reduction steps on the final calibration will be
discussed throughout the paper.

\section{Phase correction}
\label{sec:phase_correction}

Phase correction is done when the combined interferometric cube is
transformed to a spectral cube via a discrete Fourier transform
operation. It has an impact on flux and wavelength calibrations
\citep[see e.g.][]{Bell1972}. A bad phase correction will result in a
deformation of the instrument line shape (ILS) which leads to flux and
wavelength errors.

\begin{figure}
  \includegraphics[width=\linewidth]{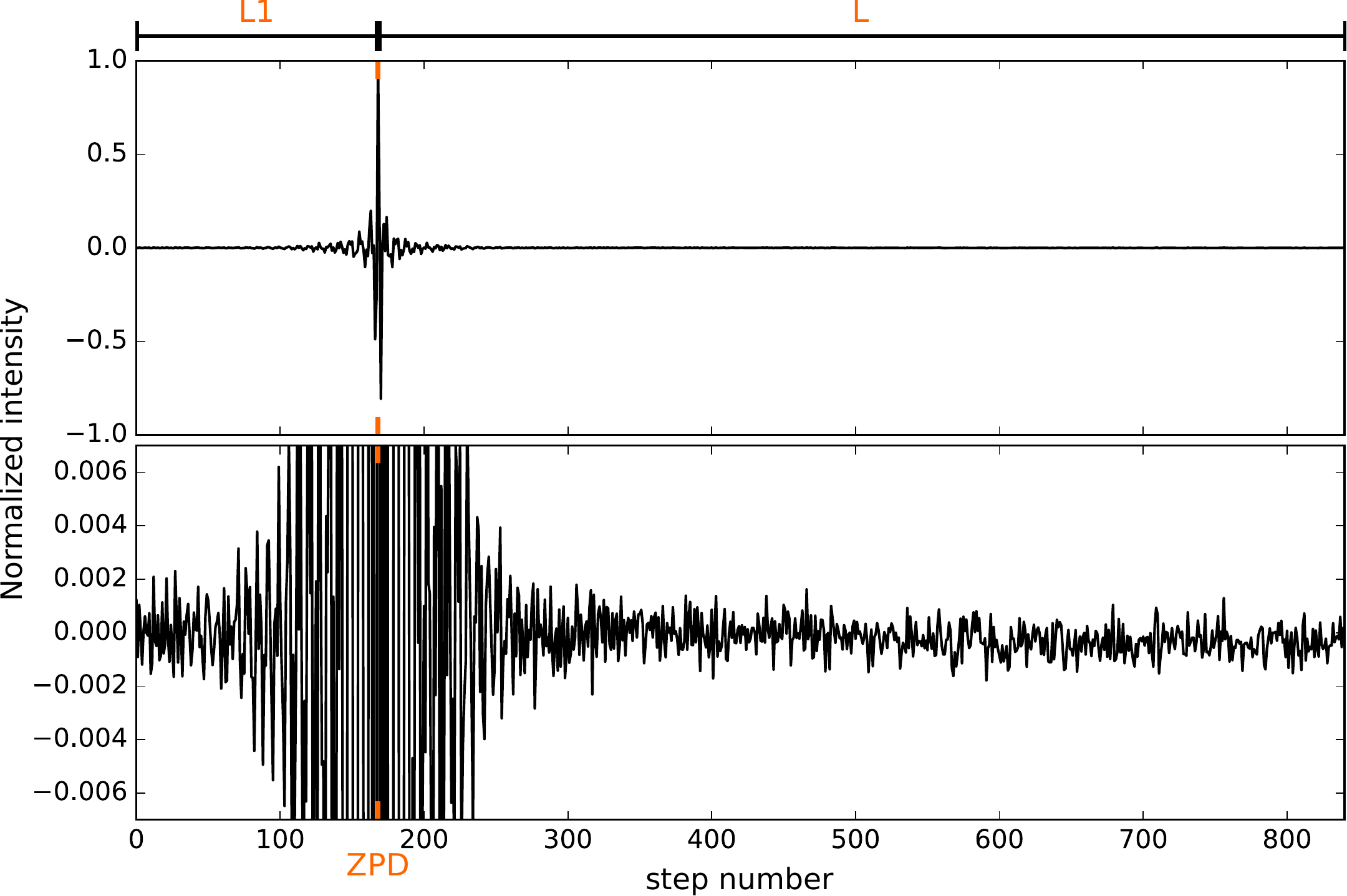}
  \caption{Example of a typical continuum source interferogram
    obtained with SITELLE. It has been recorded near the center of M31
    in the SN3 filter (courtesy of Anne-Laure Melchior). The top panel
    shows the full interferogram and the bottom panel shows a zoomed
    version. The ZPD (step 168, indicated with an orange tick) is
    positioned at the highest measured intensity of the
    interferogram. The maximum OPD attained on the left side is $L_1$
    and $L$ on the right side (see equations~\ref{eq:R_phase}
    and~\ref{eq:R_spec}.}
    \label{fig:m31-interf}
\end{figure}

If the source has a spectral distribution $S(\sigma)$, the intensity
measured at an optical path difference (OPD) $x$ will be
\citep{Sakai1968}:
\begin{equation}
  \label{eq:ft_inv}
  I(x) = \int_{-\infty}^{+\infty} S(\sigma) e^{i\phi(\sigma)}e^{2i\pi\sigma x}\;\text{d}\sigma
\end{equation}
where $\phi(\sigma)$ is a phase error which may arise from dispersive
effects in the optics of the interferometer and a non perfect
sampling. Note that, when both the fixed and the moving mirrors are
at the same distance from the beamsplitter, the OPD $x$ is equal to 0
(this position is called the Zero Path Difference, ZPD); in the
absence of phase error ($\phi(\sigma) = 0$), the intensity measured is
equal to the total intensity of the source. Which means that, at ZPD,
an interferogram is at its maximum intensity (see
figure~\ref{fig:m31-interf}).

\begin{figure}
  \includegraphics[width=\linewidth]{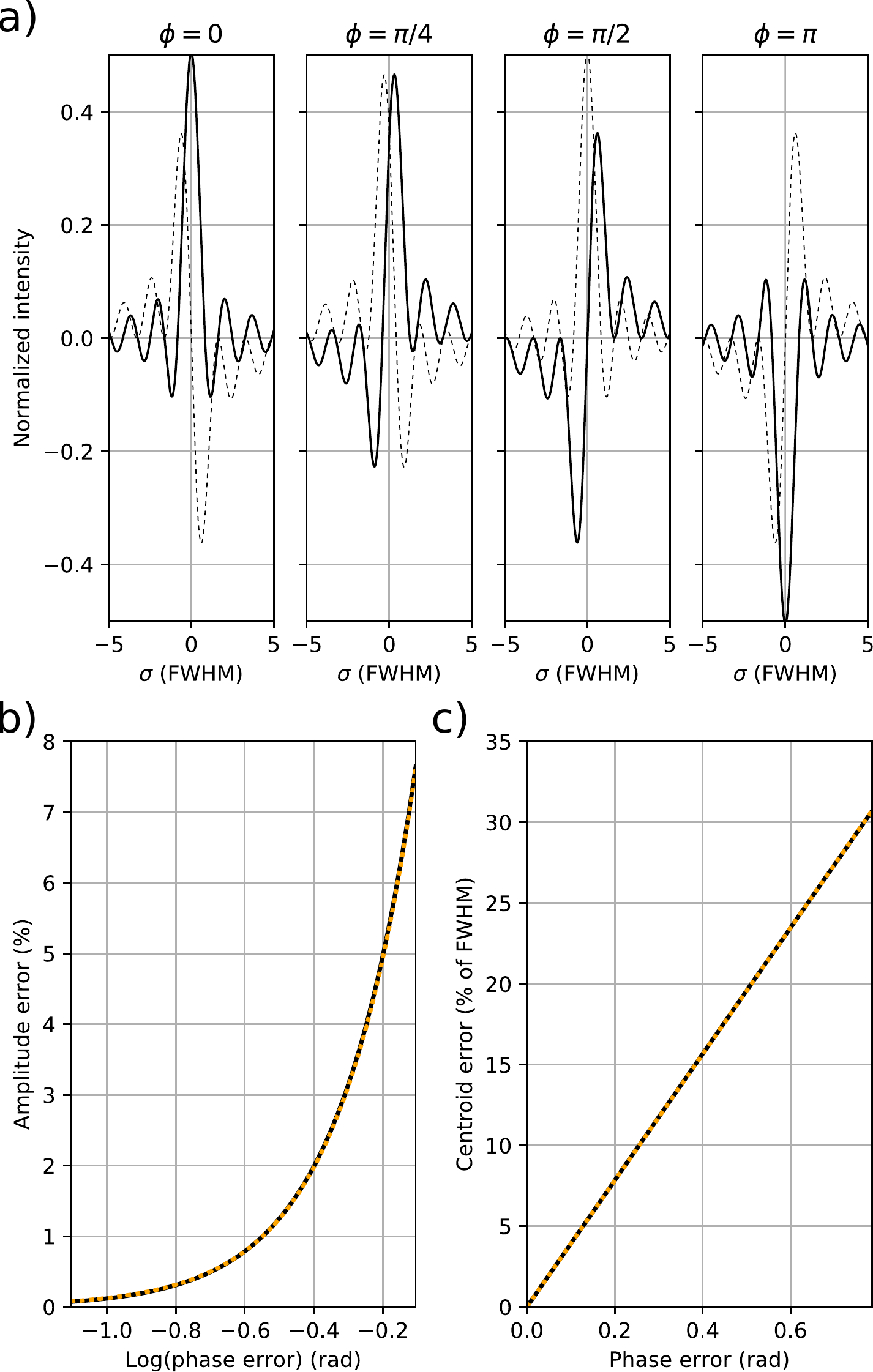}
  \caption{\textit{a)} Effect of a constant phase error on the
    instrument line shape. The real part is drawn in solid black and
    the imaginary part is dotted. The wavenumber $\sigma$ is given in
    units of the line's FWHM. \textit{b)} Effect of constant phase error
    on the amplitude of the line. The black line represents the numerical simulation and
    the dotted orange line represents the curve given in
    equation~\ref{eq:flux_error_percents}. Note that phase error
    (given in log) goes from 0.1$\pi/4$ to $\pi/4$. \textit{c)} Effect
    of constant phase error on the measured position in percentage of
    the line FWHM.}
    \label{fig:phase_err}
\end{figure}

The Fourier transform of a non-corrected interferogram is
therefore
\begin{equation}
  \label{eq:ft_cos}
  \hat{I}(\sigma) = S(\sigma)e^{i\phi(\sigma)} = \int_{-\infty}^{+\infty} I(x) e^{-2i\pi\sigma x}\;\text{d}x
\end{equation}
In the absence of phase errors, the spectrum corresponds to the real
part of the Fourier transform. However, any uncorrected phase error
will translate into an error in the calculated spectrum (see
Figure~\ref{fig:phase_err}). A first approach consists in computing
the power spectrum to remove the phase error
\begin{equation}
  \label{eq:power_spectrum}
  P(\sigma) = \left|\hat{I}(\sigma)\right|^2 = S(\sigma)^2 e^{i\phi(\sigma)}e^{-i\phi(\sigma)} = S(\sigma)^2\;.
\end{equation}
However, the ILS is modified, the spectral resolution is degraded and
the noise properties of the spectrum are much different as its
distribution is squared. A better approach to recover the real
spectrum is therefore to compute the phase and remove it from the
Fourier transform
\begin{equation}
\label{eq:phase_correction}
  S(\sigma) = \text{Re}\left(\hat{I}(\sigma)\;e^{-i\phi(\sigma)}\right)\;.
\end{equation}
The whole challenge of phase correction is thus to recover the phase
$\phi(\sigma)$, which is different from one interferogram to the
other, with the highest precision possible so that the quality of the
spectrum is ideally not affected by this operation.

\subsection{Determination of the phase}
\label{sec:estimation_of_the_phase}
The discrete Fourier transform of an interferogram of N samples is
\begin{equation}
\label{eq:fourier_transform_discrete}
  \hat{I}(\sigma) = \sum_{n=0}^N I(x_n) \exp(-2i\pi\sigma x_n) = \hat{I}_{\text{Re}}(\sigma) + i\,\hat{I}_{\text{Im}}(\sigma)\;.
\end{equation}
The phase $\phi(\sigma)$ of the complex spectrum is simply
\begin{equation}
\label{eq:phase_basic}
  \phi(\sigma) = \tan^{-1}\left(\frac{\hat{I}_{\text{Im}}(\sigma)}{\hat{I}_{\text{Re}}(\sigma)}\right)\;.
\end{equation}
If $\phi(\sigma)$ is not 0, the corrected spectrum is calculated by
removing the phase before taking the real part (see
equation~\ref{eq:phase_correction}).

In the absence of prior informations, the phase is an unknown function
of the wavenumber and can be different from one pixel to another
(i.e. different for each interferogram of the cube). It is considered
as a slowly varying function which, when computed from
equation~\ref{eq:phase_basic}, contains some noise that must be
removed before it is used to correct the spectrum. A polynomial model
is thus fitted to obtain a noiseless phase which can be interpolated
between the samples calculated with the DFT \citep{Sakai1968}.
\begin{equation}
  \label{eq:poly_phase}
  \phi(\sigma) \equiv p_0 + p_1\sigma + p_2\sigma^2 + \dots
\end{equation}

If the interferogram is asymmetric, only the symmetric part around the
ZPD is used to compute the phase. Ideally, one would try to obtain a
symmetric interferogram to maximize the quality of the calculated
phase. But, on the other hand, as the resolution depends on the
maximum path difference attained with respect to the ZPD (see
equation~\ref{eq:R_spec}), the samples recorded symmetrically
contribute to the SNR only. Therefore, in order to maximize the
attained resolution for a given exposure time, SITELLE's
interferograms are not recorded symmetrically. The shortest side of
OPD, $L_1$, before reaching the ZPD, extends to 25\% of the longest
side of OPD $L$ (see Figure~\ref{fig:m31-interf}). As the phase is
computed with the samples that are distributed symmetrically around
the ZPD \citep[e.g.][]{Bell1972}, the resolution of the phase vector,
$R_\phi$, depends on $L_1$ (see e.g. \citealt{Martin2016} for more
details on the calculation of the resolution):
\begin{equation}
  \label{eq:R_phase}
  R_\phi=\frac{2\sigma L_1}{1.20671}\;,
\end{equation}
while the resolution of the spectrum is 4 times higher,
\begin{equation}
  \label{eq:R_spec}
  R=\frac{2\sigma L}{1.20671}\;.
\end{equation}

Ideally, once the phase at each pixel is fitted with a high order
polynomial, a 3D phase cube is obtained that can be used directly to
correct the spectral cube, spectrum by spectrum, with no further
considerations. Each order of the fitted polynomial is mapped
pixel-by-pixel and these maps contain enough information to reproduce
the phase cube. In fact, the SNR of the phase vector is rarely high
enough or not homogeneous enough in the whole passband that such a
straight approach can be used. Indeed, from
equation~\ref{eq:phase_basic}, we can see that the precision of the
phase measured in a given channel depends on the SNR of the real part
and the imaginary part of the spectrum. As the noise is equally
distributed over all the channels, the precision on the phase is thus
directly proportional to the spectral energy distribution of the
observed source. An emission-line spectrum, typical of an \HII{}
region, will therefore generate a reliable phase information only at
the wavelength of the lines, while a continuum source (like a star, a
galaxy or the earth atmosphere) will be much better at providing a
reliable phase in the whole observed passband \citep{Learner1995}.

Therefore, the quality of the measured phase value will be different
from one channel to the other and also from one spectrum to the
other. The spatial correlation of the pixels, along with some
modelling, must be used to enhance the quality of the computed phase
vectors. As the model used to fit the phase of all the spectra is a
polynomial, each order of the polynomial can be mapped and each map
can be fitted with a model. Having models which depend on a small
number of meaningful parameters, such as optical and mechanical
parameters, thus appears fundamental.

\subsubsection{First order}
\label{sec:first_order_phase}
The first order of the polynomial, $p_1$, reflects the fact that the
ZPD is not properly sampled, i.e. the OPD, measured on-axis, of the
sample nearest to the real ZPD is $\epsilon_x \neq 0$ (see e.g.
\citealt{Davis2001}). The real OPD at a given step is
\begin{equation}
  x = n\delta_x  + \epsilon_x\;,
\end{equation}
with $\delta_x$ the step size and $n$ the step index (with respect to
the ZPD sample where $n=0$). The phase value in
equation~\ref{eq:ft_inv} becomes 
\begin{equation}
  \phi_1(\sigma) = 2\pi\sigma\epsilon_x\;.
\end{equation}
This relation is not complete since we know from
equation~\ref{eq:opd_theta} that the OPD depends on the incident angle
and so does the sampling shift $\epsilon_{x,\theta}$
\begin{equation}
  \label{eq:epsilon_x}
  \epsilon_{x,\theta} = \epsilon_x \cos(\theta)\;.
\end{equation}
We can now write the first order phase created by this sampling shift
at any incident angle
\begin{equation}
  \phi_1(\sigma) = p_1(\theta) = 2\pi\sigma\epsilon_{x,\theta} = 2\pi\sigma\epsilon_x \cos(\theta)\;.
\end{equation}
As the incident angle can be directly deduced from the observation of
a laser source illuminating an integrating sphere (see
section~\ref{sec:wavelength_calibration} and
equation~\ref{eq:costheta}), only one parameter, $\epsilon_x$, is
necessary to fit the map derived from the independent fit of the phase
of all the spectra of the cube.

\subsubsection{Orders larger than 1}
\label{sec:high_order_phase}
\begin{figure}
  \includegraphics[width=\linewidth]{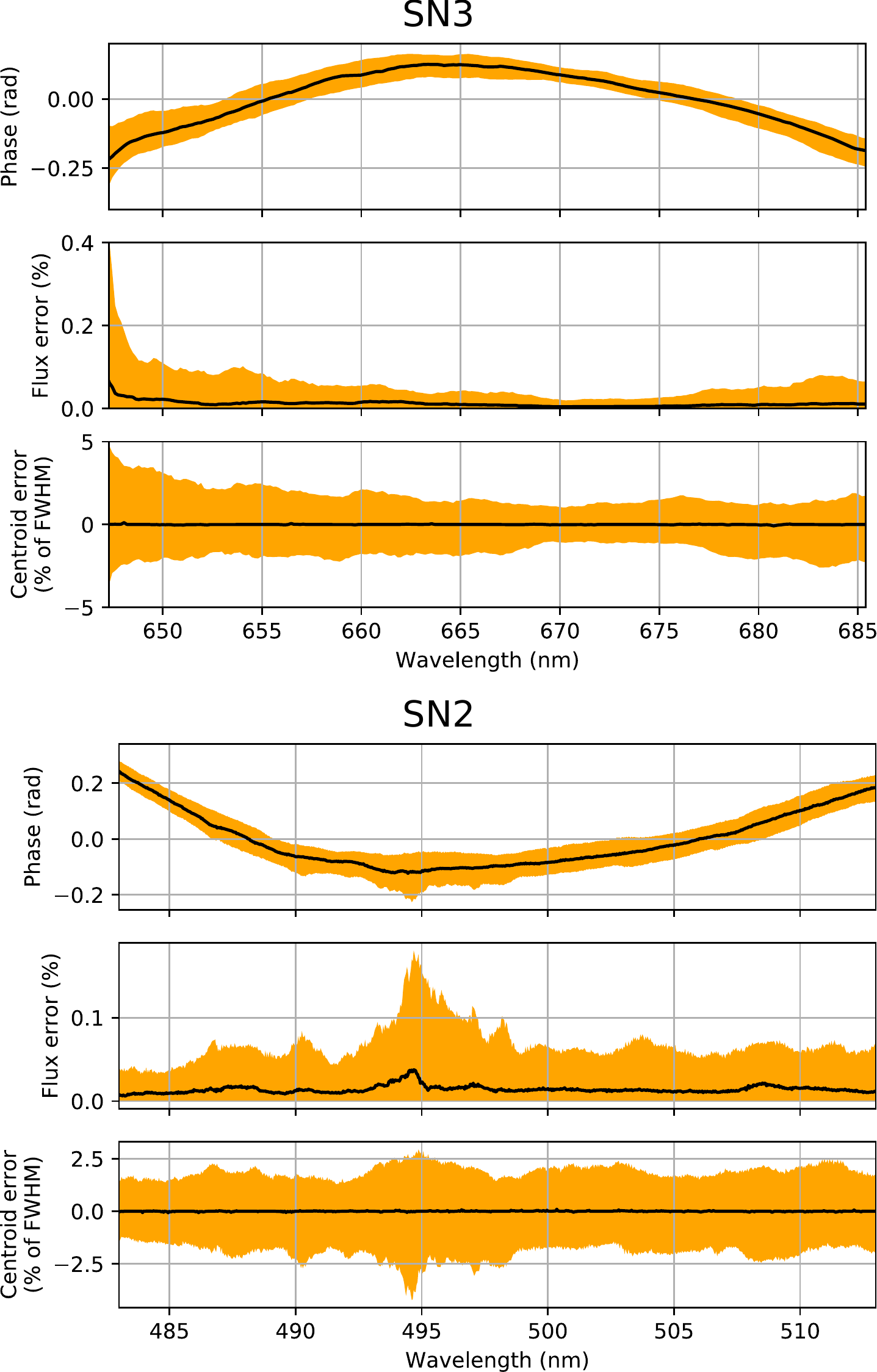}
  \caption{Median high order phase in the SN2 and SN3 filters.
    \textit{Top quadrant:} Median phase computed along the wavenumber
    axis. The values of the phase between the 16th and 84th
    percentiles are represented as an orange surface. \textit{Center
      and bottom quadrants:} Distribution of the flux error (center
    quadrant) and line centroid error (bottom quadrant) made by using the
    median phase instead of a phase computed for each pixel of the
    cube (see equations~\ref{eq:flux_error_percents}
    and~\ref{eq:centroid_error_percents}). The median error is in
    black and the 16th and 84th percentile of the error distribution
    is represented as an orange surface.}
    \label{fig:phase_high_order}
\end{figure}

All phase orders larger than 1 are considered to reflect some kind of
dispersive effect. For example, if the beamsplitter and the
compensator plate do not have the exact same thickness (see
Fig.~\ref{fig:michelson}), the phase becomes (see
e.g. \citealt{Davis2001})
\begin{equation}
  \label{eq:order1}
  \phi_{\geq 2}(\sigma) = 2\pi\sigma r(\sigma)d(\theta)\;.
\end{equation}
with $d(\theta)$ the thickness difference, which may change with
$\theta$, and $r(\sigma)$ the refractive index. For this data release
we have considered that $d(\theta)$ was constant enough that the
higher order of the phase could be approximated by the same function
everywhere in the cube.  We have checked that the error made with this
approximation was smaller than 1\% in relative flux error and that the
relative wavelength calibration error was below 5\,\% of the FWHM,
i.e. the difference between the median high order phase used for phase
correction and the high order phase measured at any point of the cube
is smaller than 0.1\,rad (see figure~\ref{fig:phase_high_order}).

\subsubsection{Zeroth order of the phase}
\label{sec:zero_order_phase}

In the ideal case, the phase shift between the two beams interfering
at the beamsplitter is $2\pi\sigma x + \pi$
\citep[p.113]{Bell1972}. We have already seen that the OPD $x$ can be
subject to a shift plus additional dispersive effects that explain
non-zero polynomial coefficients at orders larger than 0. The
constant term $\pi$ is also subject to change. For example, absorption
in the beamsplitter will change the constant value
\citep[p.125]{Bell1972} and therefore the constant phase
\begin{equation}
\label{eq:order0_nonpi}
\phi_0(\sigma) = \pi + 2\arctan{\frac{2 K(\sigma)}{r(\sigma)^2 + K(\sigma)^2 -1}}
\end{equation}
with $r(\sigma)$ and $K(\sigma)$ respectively the refractive and
absorption indexes. 

The effect of an error on the constant term can be studied by
considering an erroneous phase $\phi(\sigma) = \delta p_0$ applied to
a perfect interferogram. Equation~\ref{eq:phase_correction} can then be
written
\begin{align}
  S(\sigma) &= \text{Re}\left(\hat{I}(\sigma)\;e^{-i \delta p_0}\right)\\
  &= \hat{I}_{\text{Re}}(\sigma) \cos(\delta p_0) + \hat{I}_{\text{Im}}(\sigma)
  \sin(\delta p_0)
\end{align}
This will result in mixing the real and imaginary parts of the
spectrum (see figure~\ref{fig:phase_err}). The error made on the line
amplitude is then:
\begin{equation}
  \label{eq:flux_error_percents}
  \Delta_{\text{Flux}}\,[\%] = 100 \times \left[1 - \cos\left(\frac{\delta p_0}{2}\right)\right]\;.
\end{equation}
The error made on the line centroid is, in percentage of the FWHM,
\begin{equation}
  \label{eq:centroid_error_percents}
  \Delta_{\text{Centroid}}\,[\%] =39.1 \times \delta p_0\;.
\end{equation}
Note that these relations have been empirically calculated from a
numerical simulation. The results of the numerical simulation are
plotted in Figure~\ref{fig:phase_err}. As the phase is a slowly
varying function of the wavelength, these relations can be used to
compute the effect of a phase error no larger than $\pi/4$ at a given
wavelength (the phase error is then considered as locally
constant). As one can see on Figure~\ref{fig:phase_err}, when the
phase error is larger than $\pi/4$ the negative lobe becomes important
and even the notion of ``line'' starts to be doubtful.

\subsubsection{Phase correction method}

\begin{figure}
  \includegraphics[width=\linewidth]{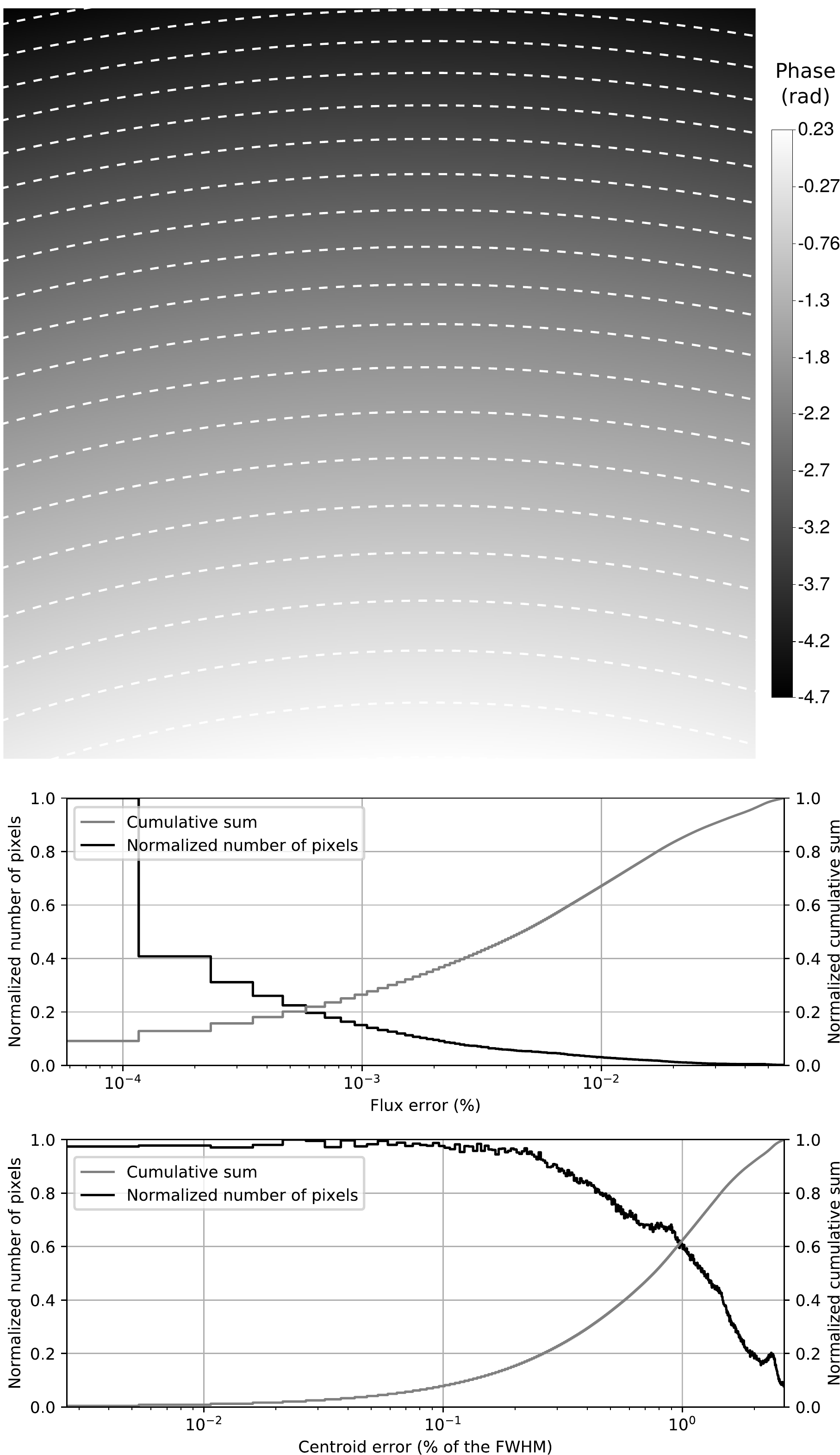}
  \caption{\textit{Top:} Typical map of the order 0 phase coefficient,
    contours have been over-plotted.  (from a scan of M31 in the SN3
    filter, courtesy of Anne-Laure Melchior). \textit{Center and
      bottom:} Histograms of the residual between the phase map and
    the fitted model in terms of flux error (center,
    equation~\ref{eq:flux_error_percents}) and line centroid error (bottom,
    equation~\ref{eq:centroid_error_percents}). The cumulative sum
    corresponding to the histogram is also shown. Both curves are
    normalized.}
    \label{fig:phase_map_order0}
\end{figure}

When the field is covered with continuum sources (earth atmosphere,
galaxies, low surface brightness diffuse gas) the two first orders of
the phase can be determined without the need for external calibration
data. In most of the cubes (galaxy clusters, star clusters,
extra-galactic objects in general), phase computation relies on the
sky background which, in most regions, is bright enough with respect
to the emission-lines of the observed astrophysical sources that a
reliable phase measurement can be obtained in all the observed band
\citep{Learner1995}. In this case a not too noisy phase vector can be
computed at each pixel of the cube, thus providing a phase cube. The
high-order phase (orders larger than 1) which has been computed from a
high-resolution scan of a white-light source observed through an
integrating sphere (see section~\ref{sec:high_order_phase}) is
subtracted and a first degree polynomial can be fitted on every phase
vector of the cube. The output from the fit are a map of order 0 which
can be fitted with the model described in
section~\ref{sec:zero_order_phase} (see
Figure~\ref{fig:phase_map_order0}) and the coefficient of order
1\footnote{In some earlier versions of the code, a map of the first
  order was computed and fitted with a spline model. In all cases the
  error made on its estimate was negligible with respect to the
  order 0.} (see section~\ref{sec:first_order_phase}). A precise phase
vector can then be reconstructed for each pixel and used for phase
correction.

\subsection{Instrument line shape}

\begin{figure}
  \includegraphics[width=\linewidth]{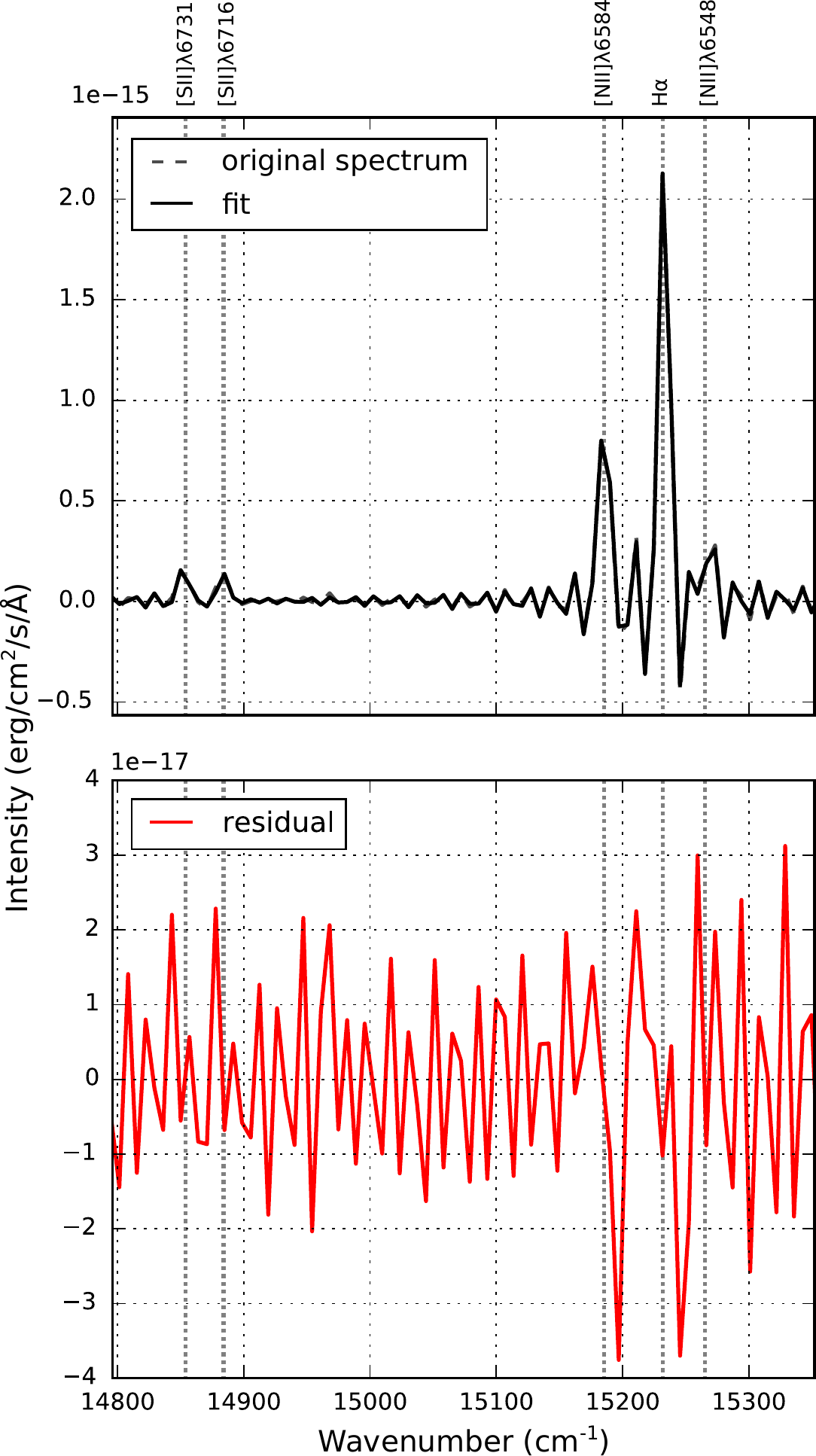}
  \caption{Example of a fit realized wit ORCS \citep{Martin2015} on a
    spectrum of a bright region of an HII region in the red SN3
    filter. The model used is the convolution of a Gaussian emission
    line and a sinc ILS \citep{Martin2016}. A modeling error is
    clearly visible near the right lobe of the emission lines.}
    \label{fig:spectrum_fit_m16}
\end{figure}

The ideal instrument line shape (ILS) of a phase corrected Fourier
transform spectrum is a sinc. Any error in the phase correction will
result in a deformation of the ILS (see
Figure~\ref{fig:phase_err}). Non-symmetric modulation efficiency loss
with the OPD (documented in \citealt{Baril2016}) can also generate an
asymmetric ILS that will eventually be the source of wavelength and
flux errors. The ILS is very well described by the theoretical model
used to fit the emission lines described in \citet{Martin2016} despite
a small modeling error that can be detected by fitting high SNR
spectra. This error is located near the right lobe of the sinc and its
amplitude, with respect to the central lobe, is around 1\% (see
Figure~\ref{fig:spectrum_fit_m16}). If this error was produced by a
phase error, a negative residual on the right lobe of the sinc would
be compensated by a positive residual on the left lobe, which is not
the case.

\section{Flux calibration}
\label{sec:flux_calibration}
\subsection{Flux calibration method}

Flux calibration is based on the measurement of the spectrum of a
spectrophotometric standard star in each filter. The obtained spectrum
is used to correct for the wavelength dependent transmission of the
instrument and the telescope. A set of images of a standard star is
also obtained at least once for each scan which provides a measurement
of the mean sky transmission in the filter band during the scan.

\subsubsection{Absolute flux calibration}

Absolute flux calibration consists in relating the real flux (in
erg\,cm$^{-2}$\,s$^{-1}$) to a measured flux of 1 count per second. As
the Fourier transform conserves the energy from the interferogram to
the spectrum (in the ideal case, when the modulation efficiency is 1),
the flux calibration calculated for an interferogram holds for the
calculated spectrum. It can therefore be made by taking a simple image
of a standard star as one would do to calibrate typical imager data.

The number $N$ of counts per second measured at one of the detectors
from a star of flux density $F_{\star}$ (in
erg\,cm$^{-2}$\,s$^{-1}$\,\AA$^{-1}$) can be written
\begin{multline}
  \label{eq:sim_counts_nb}
  N = 0.5 \int_{-\infty}^{+\infty}\frac{\lambda F_{\star}(\lambda) }{h c} T_{\text{atm.}}(\lambda) S_{\text{prim.}}(\lambda) T_{\text{tel.}}(\lambda) T_{\text{opt.}}(\lambda) \\ \times T_{\text{filter}}(\lambda) \; G  \; QE(\lambda) \,\text{d}\lambda\;,
\end{multline}
where $h$ is the Planck constant, $c$ the speed of light,
$T_{\text{atm.}}$, $T_{\text{tel.}}$, $T_{\text{opt.}}$ and
$T_{\text{filter}}$ are respectively the transmission coefficient of
the atmosphere, the telescope optics, SITELLE's optics and the filter;
$G$ is the gain and $QE$ the quantum efficiency of the CCD. The factor
0.5 comes from the fact that, because of the beamsplitter, only half
the input light is recorded by one of the
detectors. Equation~\ref{eq:sim_counts_nb} can be used to calculate the
number of counts measured for a star with a known flux
density. Typical values of the different terms for each filter are
given in Table~\ref{tab:me_filter}.
\begin{table}
  \centering
  \caption{Typical values of the transmission terms calculated for
    each SITELLE's filters. Telescope transmission is based on the
    values provided by the CFHT. The atmospheric transmission at one
    airmass comes from \citet{Buton2012}. The optics transmission have
    been provided by ABB Incorporated. Note that there are some
    differences with the previous estimations reported in the figure~9
    of \citet{Grandmont2012}.}
  \label{tab:me_filter}
  \begin{tabular}{l|ccccc}
    Filter&SN1&C1&SN2&C2&SN3\\
    \hline
    ME&0.6&0.7&0.8&0.8&0.8\\
    $T_{\text{atm.}}$&0.73&0.83&0.89&0.91&0.95\\
    $T_{\text{tel.}}$&$<$0.92&0.92&0.92&0.91&0.90\\
    $T_{\text{opt.}}$&0.67&0.69&0.67&0.69&0.66\\
    $T_{\text{filter}}$&0.94&0.97&0.97&0.95&0.97\\
    QE&0.69&0.90&0.88&0.90&0.92\\
  \end{tabular}
\end{table}

When we calibrate our data with an image the number of counts per
second is a measured quantity that must be related to the intensity of
the star in the filter passband (in erg\,cm$^{-2}$\,s$^{-1}$).
\begin{equation}
  \label{eq:standard_flux}
  \overline{F} = \int_{-\infty}^{+\infty} F_{\star}(\lambda) \tilde{T}_{\text{filter}}(\lambda) \,\text{d}\lambda \simeq \int_{\lambda_{\text{min}}}^{\lambda_{\text{max}}} F_{\star}(\lambda) \,\text{d}\lambda\;,
\end{equation}
where $\tilde{T}_{\text{filter}}$ is the normalized filter function;
$\lambda_{\text{min}}$ and $\lambda_{\text{max}}$ are the minimum and
maximum wavelength of the equivalent ideal filter bandpass.

Even if SITELLE's modulation efficiency is very high in the whole
visible band, it is never equal to 1 \citep{Drissen2010} so that the
calibration coefficient $\alpha$ by which a spectrum in
ADU\,s$^{-1}$\,\AA$^{-1}$ must be multiplied is
\begin{equation}
  \label{eq:flux_calib_relation}
  \alpha = \frac{N}{\overline{F}\times ME}\;,
\end{equation}
where $ME$, the modulation efficiency, is defined as the ratio of the
amplitude of the modulated signal over the amplitude of the input
signal. 
\begin{equation}
  \label{eq:me}
  \text{ME}=\frac{\text{Amplitude of the modulated signal}}{\text{Amplitude of the input signal}}\;.
\end{equation}
If the signal is monochromatic, it can be measured everywhere. In all
other cases, its reference value must be measured at the ZPD. The
modulation efficiency is very sensitive to the quality of the optics
(reflection index and surface quality of the mirrors and the
beamsplitter) and their alignment during the exposure
\citep[e.g.][]{Drissen2010,Maillard2013, Baril2016}.

A rough measurement of the modulation efficiency at ZPD has been made
based on simulated values by ABB and cubes of laser sources observed
at different wavelengths during the testings of the instrument before
it was shipped to the CFHT (see Table~\ref{tab:me_filter}). The
precision on this measurement is around 5\,\% and is likely to vary
from one scan to another. When the amount of stray light is
negligible, the modulation efficiency is the ratio between the energy
of the input inteferogram and the computed spectrum. As a certain
amount of stray light has been measured in SITELLE's interferometric
images, a robust algorithm still has to be developed in order to
obtain a much more precise measurement of the ME that would ideally
reflect the real mean modulation efficiency of each observation.

Note that airmass is not taken into account in the atmospheric
transmission coefficient. As its value is different for the standard
star and the science target we must correct the calibration
coefficient by the difference of transmission due to the
airmass. However, this correction has not been considered for this
data release as its impact is negligible with respect to the other
sources of uncertainty (see section~\ref{sec:influence_on_flux}).

\subsubsection{Wavelength dependent flux calibration}

\paragraph{Filter transmission}

\begin{figure}
 \centering
 \includegraphics[width=\linewidth]{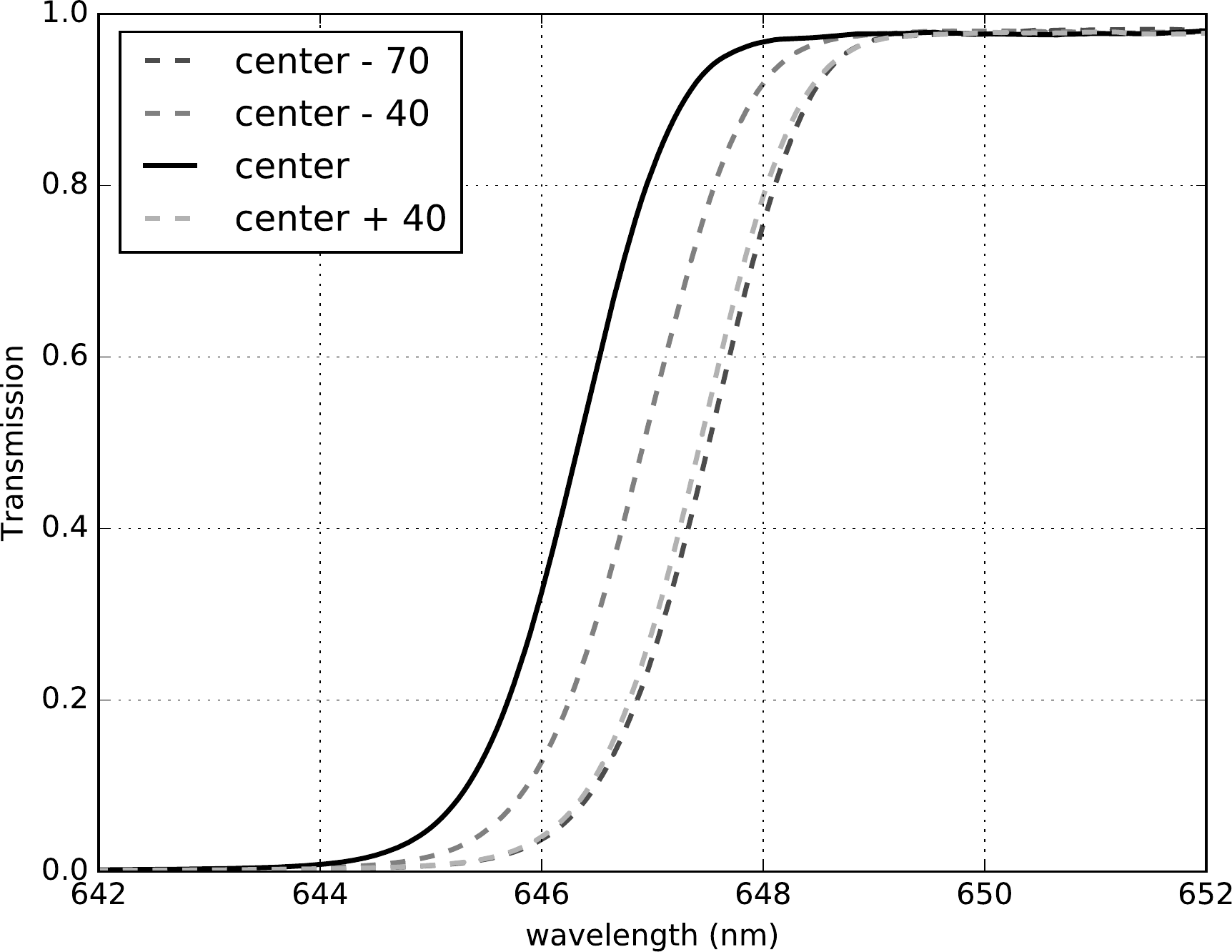}
 \caption{Edge of the SN3 filter at different positions in the field
   of view.}
 \label{fig:filter_sn3}
\end{figure}

\begin{figure}
  \includegraphics[width=\columnwidth]{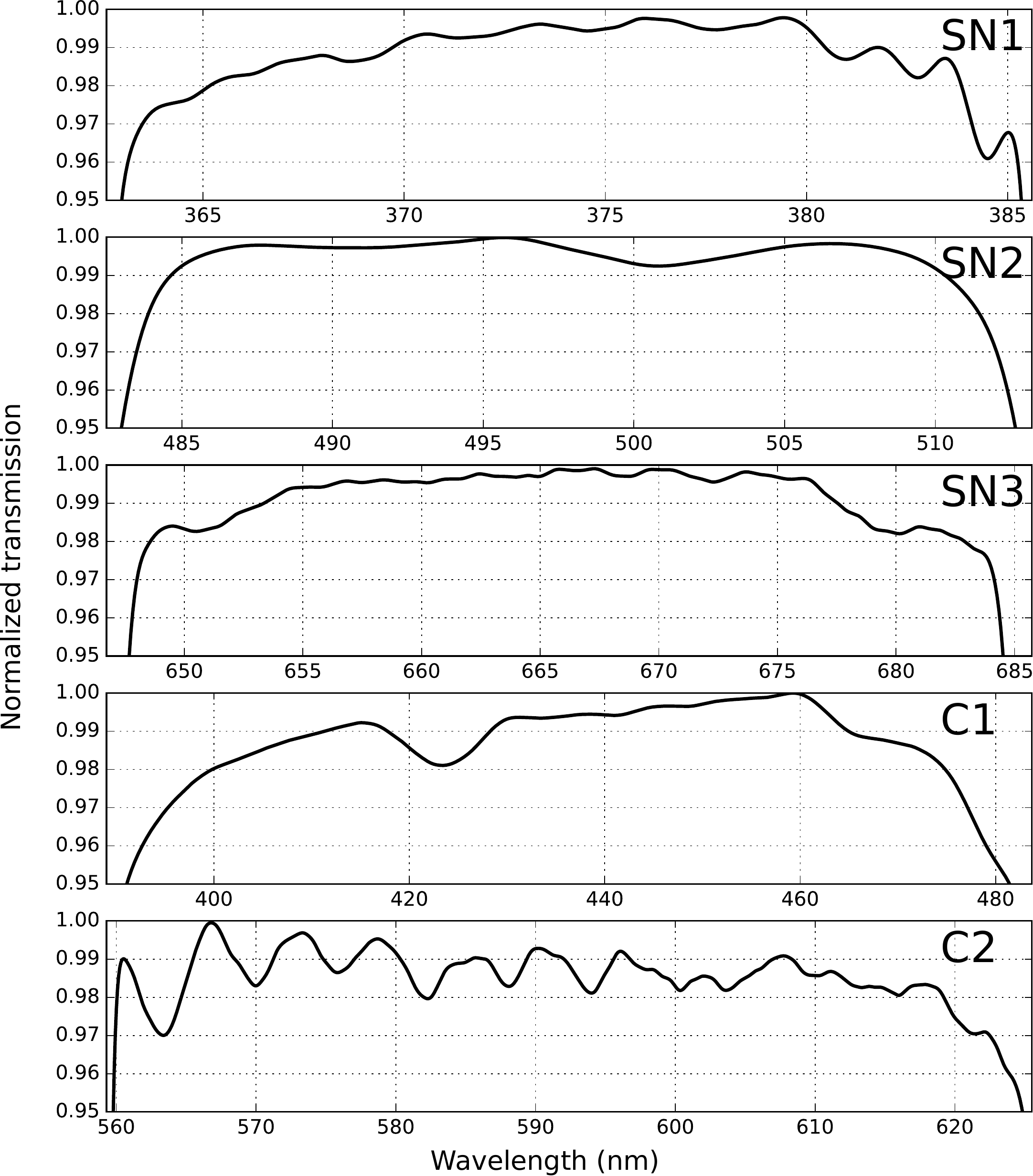}
  \caption{Normalized SITELLE's filter transmission curves. Note that
    filters SN2 and C1 have been scanned at a lower resolution and may
    seem falsely smoother than the others.}
    \label{fig:curves}
\end{figure}

As the filter passband is not the same everywhere in the field (see
Figure~\ref{fig:filter_sn3}), filter correction requires the
determination of a 3D transmission function which was not known
precisely enough for the data release 1. Therefore, no filter
correction has been done. Note that the edges of the filters are quite
steep and the peak-to-peak difference in the passband is not higher
than 4\% and generally smaller than 2\% (see Figure~\ref{fig:curves}).


\paragraph{Instrumental transmission}

\begin{figure}
  \includegraphics[width=\columnwidth]{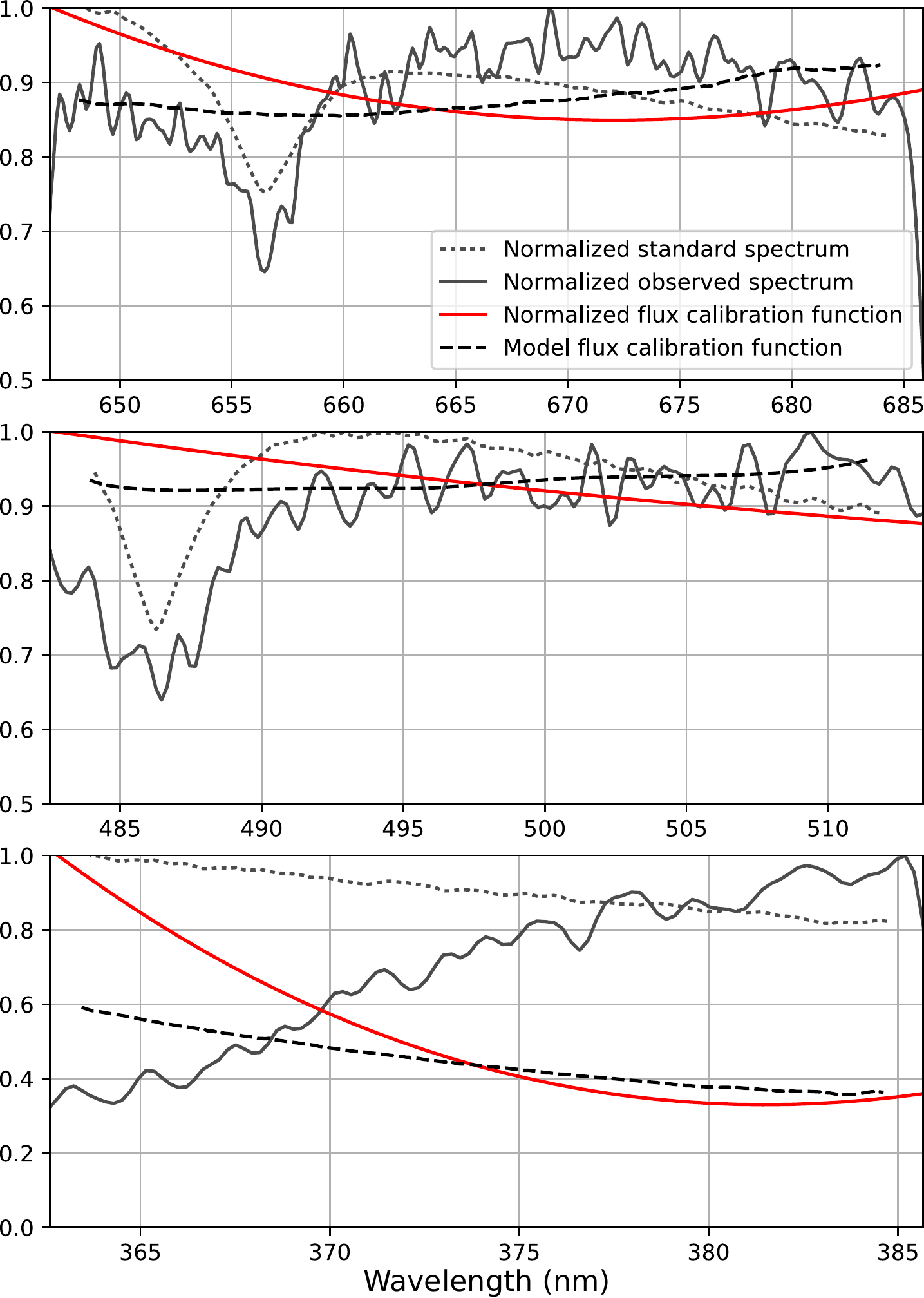}
  \caption{Normalized SITELLE's flux calibration functions. The
    observed spectrum of the standard star and the standard spectrum
    are shown. The model calibration function based on our knowledge
    of the different transmission functions considered in
    equation~\ref{eq:t_instrument} is also shown. This model is not
    used for the calibration.}
    \label{fig:fcalib_all}
\end{figure}

Following equation~\ref{eq:sim_counts_nb}, we see that a lot of
optical media with different transmission functions are contributing
to the whole instrument transmission (here the concept of
``instrument'' is equivalent to the combination of SITELLE, the
telescope and the atmosphere). The knowledge of the exact impact of
each parameter of this function is unimportant since we can determine
the instrumental transmission $T_{\text{instrument}}$ from the
measured spectrum of a standard star. Indeed, we can write
\begin{multline}
  \label{eq:t_instrument}
  T_{\text{instrument}}(\lambda) = T_{\text{atm.}}(\lambda)\, T_{\text{tel.}}(\lambda)\, T_{\text{opt.}}(\lambda)\, \\\times T_{\text{filter}}(\lambda)\; G \; QE(\lambda) \;,
\end{multline}
and,
\begin{equation}
  \label{eq:flux_calib_wave}
  T_{\text{instrument}}(\lambda) = \frac{\alpha S_{\star}(\lambda)}{F_{\star}(\lambda)}\;,
\end{equation}
where $S_{\star}$ is the measured spectrum in
ADU\,s$^{-1}$\,\AA$^{-1}$ and $\alpha$ is the calibration coefficient
computed from standard images (see
equation~\ref{eq:flux_calib_relation}). The calibration function by
which a spectrum must be multiplied is the inverse of the instrumental
transmission. The flux calibration functions computed for the filters
SN1, SN2 and SN3 are shown on Figure~\ref{fig:fcalib_all} along with a
model based on our knowledge of the different transmission functions
considered in equation~\ref{eq:t_instrument}. We can see serious
differences between them that must be due to our poor knowledge of the
transmission function of the telescope mirrors (especially in the blue
part of the SN1 filter). It is also clear that the important noise on
the measured standard spectrum makes the uncertainty on the relative
flux calibration around 5\,\%. Noise is very difficult to reduce when
observing bright stars because of the detectors saturation. The only
method that would seriously enhance the precision of the flux
calibration function would be to observe at the same time a very high
number of secondary calibration sources like star clusters for which
standard spectra would have been obtained independently.

\subsection{Influence of the reduction steps on the flux calibration}
\label{sec:influence_on_flux}
Flux calibration is certainly the most sensitive to the quality of
each reduction step. We are going to analyze the relative contribution
of each reduction step on its precision.

\subsubsection{Image correction and alignment}
\label{sec:image_correction}

As with any imager, the number of counts recorded in each pixel must
first be corrected for electronic and thermal biases, coming from the
detector, as well as differences in the illumination pattern due to the
whole optical system comprising the telescope and the instrument
optics. This corrected flux must also be conserved through the
alignment steps which are all based on linear interpolation. All those
first steps have an influence on the homogeneity of the measured flux
in each image taken independently.

\paragraph{Flat-field}

Flat-field correction is the most important contributor in terms of
relative flux calibration errors. Flat-field images are obtained by
observing the sky at twilight through the same filter used to observe
the science data. Five images are taken at the beginning of each night
and combined to make a master flat-field image which is used to
correct the illumination pattern of the science images. The
combination is an implementation of the averaged sigma-clipping
algorithm of the task \texttt{imcombine} of the package
\texttt{stsdas} of the IRAF pipeline \citep{TodyDoug1993}. In the
worst cases a gradient in the sky level of up to 5\% has be seen in
the corrected images. The maximum difference is seen in the corners of
the image.

\paragraph{Alignment}

Once the alignment parameters are found by measuring the positions of
the stars in the field of view with a Gaussian profile fitting
algorithm, images are geometrically transformed with a classic linear
interpolation algorithm. Even if linear interpolation will certainly
modify the shape of a star and therefore cannot guarantee flux
conservation, a simulation of this effect shows that the flux is
perfectly well conserved for stars if aperture photometry is used,
i.e. the total number of counts in a circular aperture around the star
is conserved. This result is obvious since the interpolation of any
signal will conserve the value of the integral of the signal. We have
therefore not detected any error on the flux measurement larger than
the numerical error with aperture photometry. But profile fitting will
give a worse estimate of the flux since the overall shape is
modified. For example, the peak value of the transformed point spread
function (PSF) is around 0.5 the peak value of the real PSF. If a
fitting procedure is used (in the idealized case of a Gaussian PSF),
the median error made on the flux measurement is around -2\% (flux is
always underestimated) and the largest error is always smaller than
-5\% (over 5000 randomized tries).

\subsubsection{Cube combination}
\label{sec:cube_combination}
During the combination step it is possible to correct for temporal
variations of the atmospheric transmission. Remember that the cubes
are complementary by definition, the light that does not go through
one of the output ports must go through the other. Combining both
cubes has therefore two advantages. (1) It provides two times more
photons, which enhances the signal-to-noise ratio (SNR) by a factor of
$\sqrt{2}$. (2) Given the fact that all the input light ends on both
output ports it is possible to follow the flux variation of the input
light. With the assumption that the source is not variable and that
the overall transmission remains stable during the observation, the
variation of the sum of the flux measured on both ports gives a robust
estimate of the variation of the mean sky transmission in the
observed passband (atmospheric extinction and airmass).

\begin{figure}
 \centering
 \includegraphics[width=\linewidth]{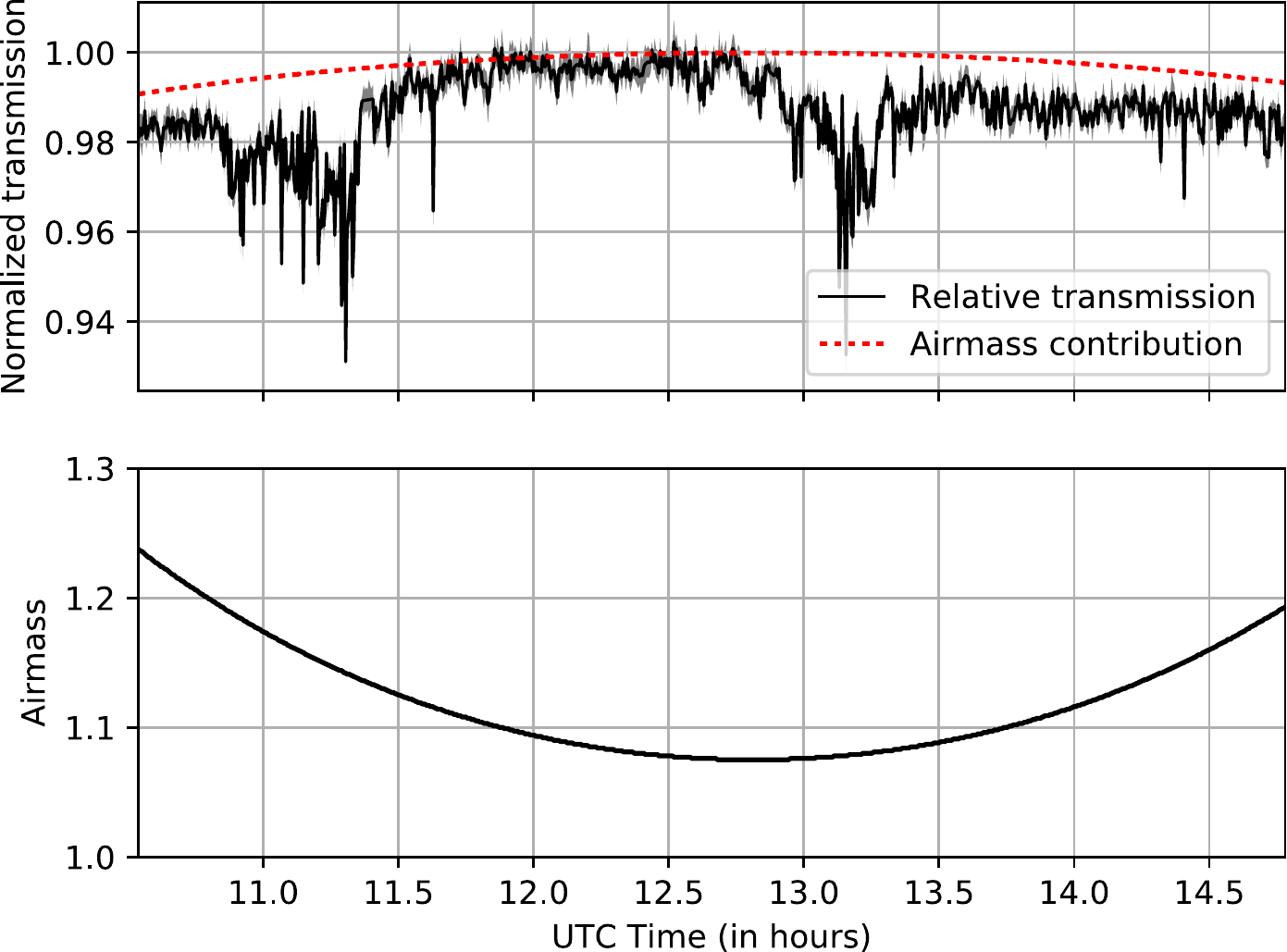}      
 \caption{\textit{Top:} Atmospheric transmission function during the
   acquisition of the M\,31 cube in the SN3 filter observed on August
   24, 2016 (courtesy of Anne-Laure Melchior, see Martin et al. 2017
   in preparation). The computed transmission is normalized to its
   99th percentile. The grey surface represents the uncertainty. The
   airmass contribution to the atmospheric transmission is plotted in
   dotted red. It as been computed with a mean value of the extinction
   over the Mauna Kea at the H$\alpha$ wavelength of
   6.2\,10$^{-2}$\,mag/am quoted from
   \citet{Buton2012}. \textit{Bottom:} Estimated airmass of the
   observed target.}
  \label{fig:atm_trans}
\end{figure}

A first order combination equation of the two interferograms of the
same source is
\begin{equation}
  \label{eq:merging_eq}
  I(t) = \frac{I_1(t) - I_2(t)}{I_1(t) + I_2(t)}\;,
\end{equation}
$I_1$ and $I_2$ being the interferograms recorded in camera~1 and
camera~2 respectively \citep{Davis2001, Martin2012} ; $t$ is the time
when each sample is recorded. Because the interferograms measured on
each arm are complementary (i.e. all the incoming flux is separated
between each arm), $I_1 - I_2$ represent the modulated part of the
interferogram while $I_1 + I_2$ is the intensity at the input of the
interferometer. The relative variation of the atmospheric transparency
is thus corrected by the denominator. As $I_1 + I_2$ must be the same
for all the pixels of the cube we can use the stars to compute an
atmospheric transmission function
$T_{\text{atm.}}(t) = I_1(t) + I_2(t)$ for all the interferograms. We
use the stars to compute the transmission because the measure of
their flux is much less sensitive to scattered light. A typical
atmospheric transmission function is shown in
Figure~\ref{fig:atm_trans}.
Equation~\ref{eq:merging_eq} can thus be rewritten
\begin{equation}
  \label{eq:merging_eq_2}
  I(t) = \frac{I_1(t) - I_2(t)}{T_{\text{atm.}}(t)}\;,
\end{equation}

There are two underlying assumptions in this equation:
\begin{itemize}
\item The detectors are sufficiently similar that the quantum
  efficiency (QE) is the same at all wavelengths.
\item Since the subtraction at the numerator will remove any amount of
  stray light equally found in both cameras, the difference in the
  amount of stray light ending in both cameras must be negligible.
\end{itemize}
While these two assumptions could not hold in the case of its
predecessor SpIOMM \citep{Grandmont2003, Bernier2006}, they are
considered true in the case of SITELLE.

\subsubsection{Phase correction}

Remember that the phase vector used to correct a given spectrum is
computed from two polynomial coefficients: order 0, which is mapped
and order 1, which reduces to one number for the whole cube and
the higher order phase, which is a function of the wavenumber
calculated from a continuum calibration source observed once by run (a
SITELLE run never lasts more than 10 nights). The spectrum
attached to pixel $k$ will be corrected with a phase vector
$\phi(k, \sigma)$:
\begin{equation}
  \phi(k, \sigma) = p_0(k) + \sigma p_1 + \phi_{\geq 2}(\sigma)
\end{equation}
The error (in percentage) on the measure of the flux resulting from
an error in the determination of the phase at any given wavenumber can be
computed from equation~\ref{eq:flux_error_percents}. If we assume that
the order 0 phase map is a slowly varying function of the position in
the field-of-view, i.e. there is no discontinuity between one pixel
and its neighbours (see Figure~\ref{fig:phase_map_order0}), then a
model can be fitted and the residual which will contain noise and
modelling errors will provide a conservative estimate of the error made
(i.e. noise is considered negligible, which is certainly not
true). Even in this case, the flux error based on this residual is
always much smaller than 1 percent and appears negligible when
compared to the flat field uncertainty.

\subsection{Assessment of the uncertainty}

\subsubsection{Uncertainty on the absolute flux calibration}

From the analysis conducted in this section we conclude that there are
two major sources of uncertainty on the absolute flux calibration: the
modulation efficiency measurement and the determination of the mean
atmospheric transmission loss from standard star images. Potential
phase correction errors seem completely negligible at this point.

\begin{figure}
 \centering
 \includegraphics[width=\linewidth]{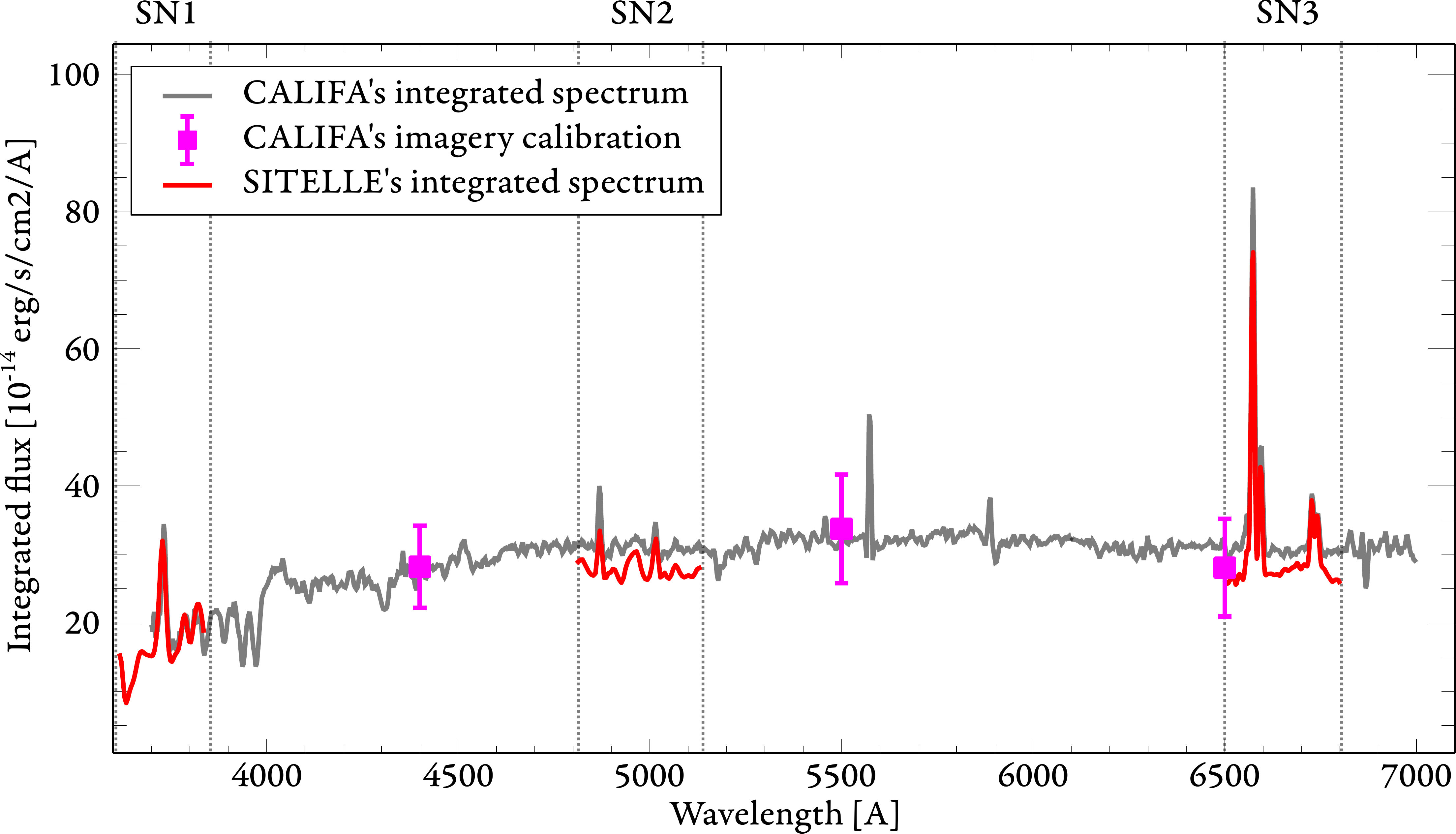}      
 \caption{Integrated spectrum of NGC\,628 obtained with SITELLE in
   three filters (SN1, SN2 and SN3) superimposed on the integrated
   spectra obtained with PPaK \citep{Sanchez2010, Kelz2006}. SITELLE's
   spectra have been convoluted to respect PPaK's low resolution. A
   correction factor of 0.65 has been applied to consider PPaK's
   filling factor. The photometric calibration points used to
   calibrate PPAK spectrum are shown in purple along with their
   uncertainty. Part of the figure has been taken from
   \citet{Sanchez2010}}
  \label{martin1:califa}
\end{figure}

We have checked the accuracy of the calibration against various
references: independent point-like sources (galaxies at z$\sim$0.3 in
the HETDEX Field, the compact planetary nebula M1-71) and the
integrated spectrum of a galaxy covering the whole field of view in
three different filters (NGC\,628, see
Figure~\ref{martin1:califa}). All the results are reported in
Table~\ref{martin1:tab_flux}. There is an obvious general bias around
-5\,\% which must come from the rough knowledge of the modulation
efficiency. A better estimate of the modulation efficiency could be
derived from the ratio of the total spectral energy present in the
output spectra and the total energy deposited by the photons in the
input interferograms. It will be corrected in a future release of the
reduction code. We also must mention that a modulation efficiency loss
of 10\,\% has been measured during the observation of the SN3 cube of
M31 (see Martin et al. in preparation). In general the modulation
efficiency does not vary much from one observation to another which
implies that, even if the estimate is biased, its precision is better
than 5\,\%. However, as for the first data release, a conservative
evaluation of the uncertainty on the modulation efficiency should be
between -10\,\% and 0\,\%. Laser frames are obtained at the beginning
and the end of each scan which should permit to calculate the
modulation efficiency loss in future releases.

As for the measure of the mean atmospheric transmission during the
scan, we have also observed that, in rare cases, the standard star
images were not taken right before or after the scan and sometimes a
night later. In these cases, the measurement is completely unreliable
and should not be trusted. The chosen method also suffers from the
fact that the transmission can vary by more than 10\,\% on a time
scale of a few minutes, which means that the atmospheric transmission
measured right after the scan sequence may not reflect precisely even
the last minutes of the observation.

An evaluation of the precision of the flux calibration, taking the 5\,\%
bias into account, is thus generally between -10\,\% and 0\,\% if we
consider the calibration checks reported in
Table~\ref{martin1:tab_flux}. But a more conservative estimation should
be considered to lie between -15\,\% and 5\,\%.

In a general case we recommend that the absolute flux calibration be
checked against external data.

\subsubsection{Uncertainty on the relative flux calibration}
\begin{figure}
 \centering
 \includegraphics[width=\linewidth]{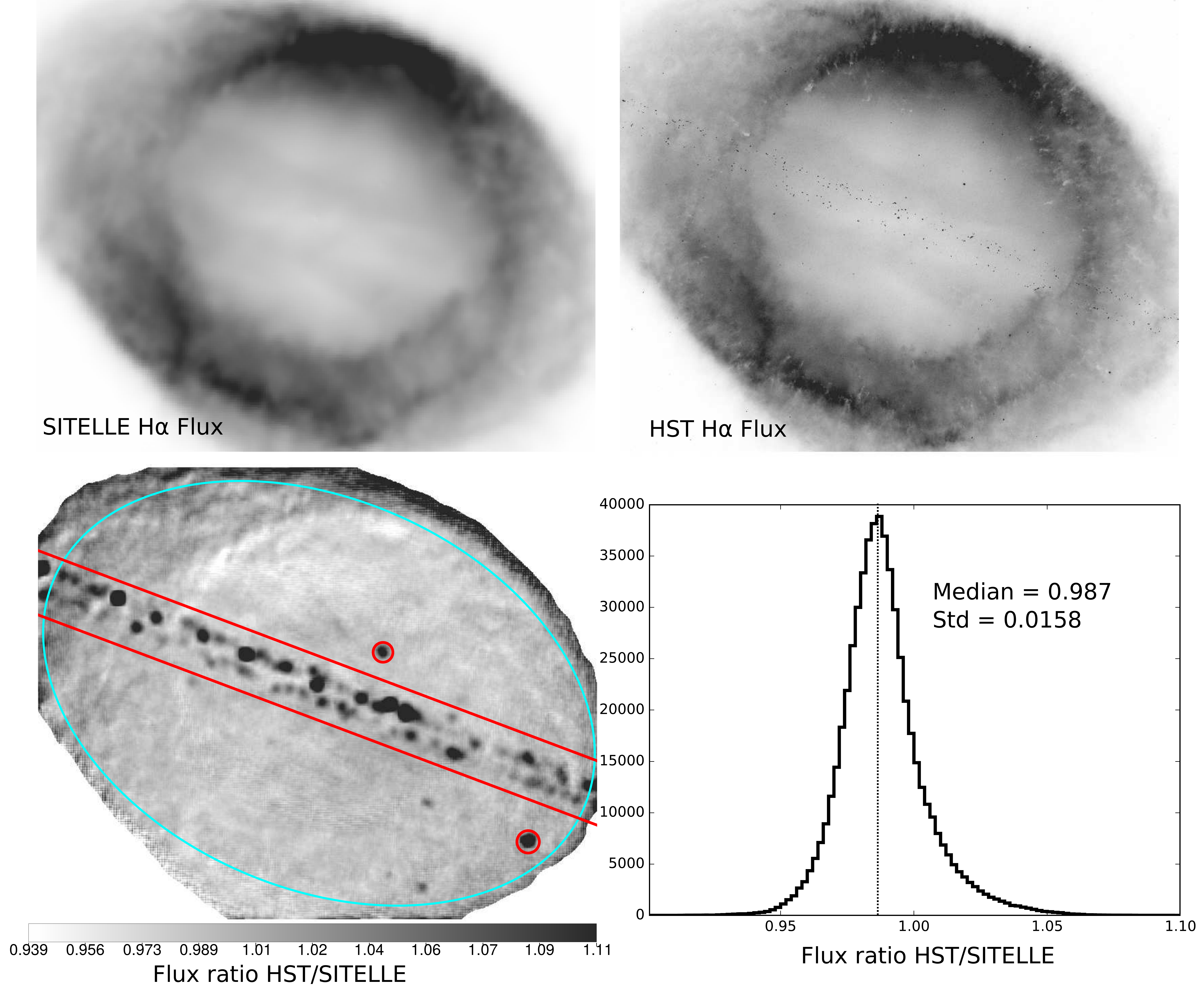}
 \caption{Comparison of the H$\alpha$ flux maps of the planetary
   nebula M\,57 obtained with SITELLE (top-left) and the image
   obtained through the F656N filter with the Hubble Space Telescope
   \citep[][top-right]{Odell2013}. The bottom-left quadrant shows the
   pixel-to-pixel flux ratio and the bottom-right quadrant shows the
   histogram of the ratios. The HST map has been smoothed with a
   8$\times$8 Gaussian kernel to fit SITELLE's pixel scale. The
   regions shown in red have been excluded from the histogram because
   they are strong stars and reconstruction errors in the HST
   mosaic. The region included in the histogram is indicated as a blue
   ellipse.}
  \label{martin1:m57-comp}
\end{figure}

The pixel-to-pixel precision of the flux calibration has been checked
by comparing the H$\alpha$ map of the planetary nebula M\,57 obtained
with SITELLE and the map obtained through the F656N filter of the
Hubble Space Telescope \citep{Odell2013}. After a careful alignment
and convolution of the HST map to respect SITELLE's pixel scale, an
histogram of the flux ratio has been computed (see
Figure~\ref{martin1:m57-comp}). We can see that the error is smaller
than 1.5\% and the standard deviation of the ratios is smaller than
1.6\%. Note that the object covers only a small part of the field of
view (around 1$\times$1 arcminute) so that large gradient cannot be
detected. Another example of the pixel-to-pixel flatness of the flux
calibration is given in Martin et al. (in preparation) for the
calibration of M\,31. Up to now, no overall gradient has been detected
and the relative flux calibration seems better than 3\,\%.

\begin{table}
  \centering
  \caption{Flux calibration check against various references. Three different filters have been checked: SN1 (362.6--385.6\,nm), SN2 (482--513\,nm) and SN3 (647.3--685.4\,nm).}
  \label{martin1:tab_flux}
  \begin{tabular}{llc}
    Object&Wavelength range&Error\\
    \hline
    NGC3344 & H$\alpha$ vs. SpIOMM&\textbf{-4\%} $\pm$2\%\\
    & H$\alpha$ + [NII]$\lambda6584$&\textbf{-4\%} $\pm$3\%\\
    \multicolumn{3}{l}{\citet{RNepton2017t}}\\
    \hline
    M1-71 & H$\alpha$&\textbf{-7\%} $\pm$3\%\\
    vs. & [NII]$\lambda6584$ &\textbf{-11\%} $\pm$3\%\\
    \multicolumn{3}{l}{\citet{Wright2005}}\\
    \hline
    NGC628&SN1&\textbf{-6\%} $\pm$6\%\\
    &SN2&\textbf{-7\%} $\pm$6\%\\
    &SN3&\textbf{-9\%} $\pm$6\%\\
    \multicolumn{3}{l}{\citet{Sanchez2010}}\\
    
    \hline
    HETDEX field&SN2 (Ly$\alpha$ flux  of  $\sim 20$ high- &\textbf{-5\%}$\pm$7\%\\
    (Drissen et al.)&redshift galaxies)&\\
    \multicolumn{3}{l}{\citet{Hill2008}}\\

  \end{tabular}
\end{table}

\section{Wavelength calibration}
\label{sec:wavelength_calibration}
Recording an interferometric image implies to measure the flux at
different angles $\theta$ with respect to the interferometer axis
where $\theta=0$. At each step of the scan, for an off-axis pixel, at
an angle $\theta$, the OPD $x_{\theta}$ with respect to the OPD
on-axis $x$ is simply \citep[see e.g.][]{Martin2016}
\begin{equation}
\label{eq:opd_theta}
  x_{\theta} = x \cos(\theta)
\end{equation}
If we consider a constant optical step size $\delta_x$\footnote{In
  this paper only optical distances will be considered. The optical
  distance along the interferometer axis $x$ is roughly 2 times the
  mechanical position of the scanning mirror. Note also that all
  distances are always measured with respect to the ZPD.}, at a given
step index $j$, the on-axis OPD is
\begin{equation}
  x = j \delta_x
\end{equation}

Fourier theorem stands that the maximum wavenumber
$\sigma_{\text{max}}$ that can be measured is inversely proportional
to two times the sampling step size.
\begin{equation}
  \sigma_{\text{max}, \theta} = \frac{1}{2\delta_x \cos(\theta)}
\end{equation}
Like its predecessor, SpIOMM, SITELLE's observing mode makes use of
spectral folding \citep{Grandmont2006, Drissen2010,
  Grandmont2012}. Because the light is observed though a filter, one
can discriminate between all the multiples of a given wavelength that
do not fall into the observed passband. It is therefore possible to
scan with a sampling step that is a multiple of the minimum observed
wavelength and increase the resolution without folding the spectral
information. The folding order $n$ of the observation can be related
to the number of times the step size is increased:
\begin{equation}
  \delta_{x,n}=(n+1)\delta_x\;.
\end{equation}
The maximum observable wavenumber, $\sigma_{\text{max}, \theta}$,  is then,
\begin{equation}
  \sigma_{\text{max}, \theta} = \frac{n+1}{2\delta_x \cos(\theta)}\;,
\end{equation}
If we want to make sure that all the wavelengths of the observed light
are discriminated by the Fourier transform we must also limit the
passband to a minimum wavenumber, $\sigma_{\text{min}, \theta}$,
\begin{equation}
  \sigma_{\text{min}, \theta} = \frac{n}{2\delta_x \cos(\theta)}
\end{equation}

\begin{figure}
 \centering
 \includegraphics[width=\linewidth]{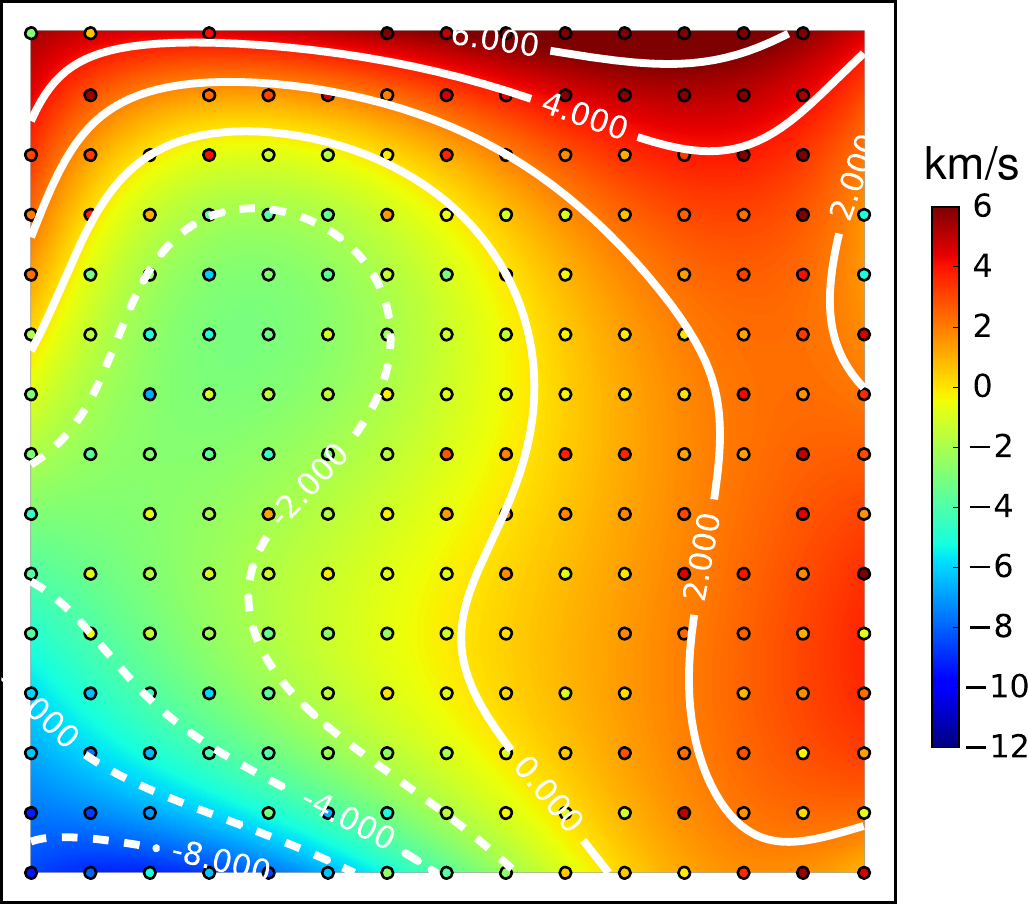}
 \caption{Relative velocity map calculated from sky lines. Extracted
   from a SN3 cube of PG1216+069 at R=1900 (courtesy of Wei-Hao
   Wang).}
  \label{martin1:velmap}
\end{figure}

\begin{figure}
 \centering
 \includegraphics[width=\linewidth]{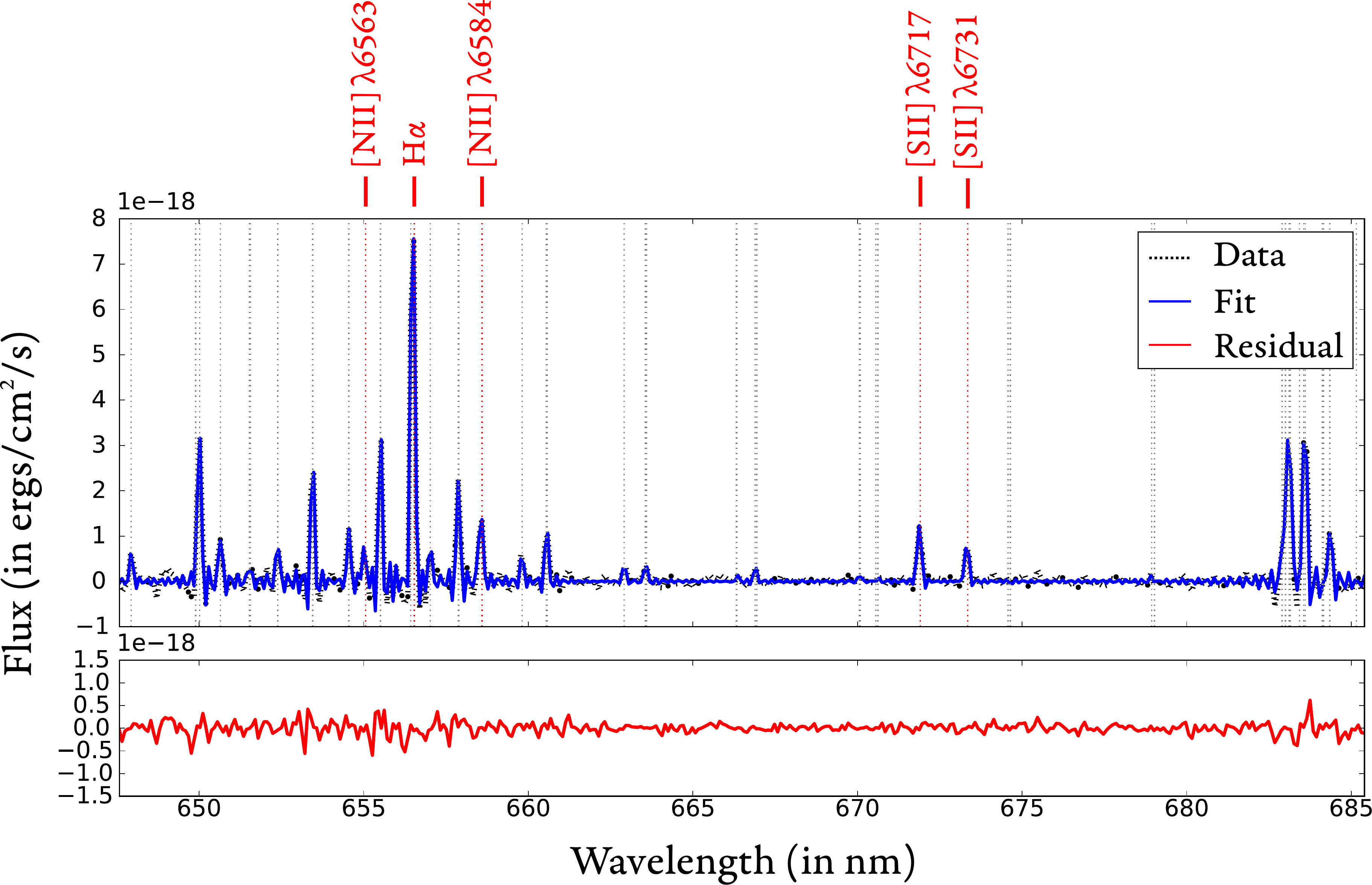}
 \caption{Example of a fit of the Meinel OH bands of a sky spectrum in
   the field of IC\,348. R = 4500 (courtesy of Gregory Herczeg). The
   fitted emission lines of the diffuse gas around the nebula are
   shown.}
  \label{martin1:skyspec}
\end{figure}

If the spectrum has $N$ samples, the wavenumber
associated to a channel $i$ of the output spectrum is
\begin{equation}
\label{eq:sigma_channel_theta}
  \sigma_{i, \theta} = \sigma_{\text{min}, \theta} + \frac{i}{N} (\sigma_{\text{max}, \theta} - \sigma_{\text{min}, \theta}) =  \frac{1}{2\delta_x \cos(\theta)} \left(n + \frac{i}{N}\right)
\end{equation}
We see that, to make an absolute calibration, the value of the
incident angle of the light for each pixel of the cube is the only
quantity needed. Inversely, if ones measures the exact position, in
channels, of the centroid of a line with a known wavelength, the
incident angle of the spectrum can be derived from
equation~\ref{eq:sigma_channel_theta}. From
equation~\ref{eq:sigma_channel_theta} we can relate directly the
wavenumber, $\sigma$, of source measured at an angle $\theta$ with
respect to the axis of the interferometer to to its real wavenumber
$\sigma_0$
\begin{equation}
\label{eq:costheta}
  \sigma = \sigma_0\cos(\theta)\;.
\end{equation}
The zero point must therefore be calibrated for each spectrum of the
cube via the observation of a laser source at zenith. One source of
uncertainty comes from the fact that the deformation of the optical
structure when the telescope moves from the Zenith position to the
direction of the source has a strong impact on the incident angle seen
by one pixel. By comparing calibration maps taken at 47 degrees in 4
directions (north, south, east, west) we have found that this
calibration method is likely to produce a gradient in the relative
wavelength calibration of up to 15\,\kms{}. Note that the
original wavelength calibration of a cube (especially in the SN3 red
filter) can be improved to a precision of a few\,\kms{} by
fitting the Meinel OH bands which are generally present everywhere in
the cube (see Figure~\ref{martin1:velmap} and
Figure~\ref{martin1:skyspec}). This operation can be done with ORCS
\citep{Martin2015} and has been used by \citet{Martin2016} and
\citet{Shara2016}. A calibration laser map model has also been
developed to increase the precision of the calibration and reconstruct
the velocity field in regions where the OH lines are not visible (see
Martin et al., in preparation).

\begin{figure}
 \centering
 \includegraphics[width=\linewidth]{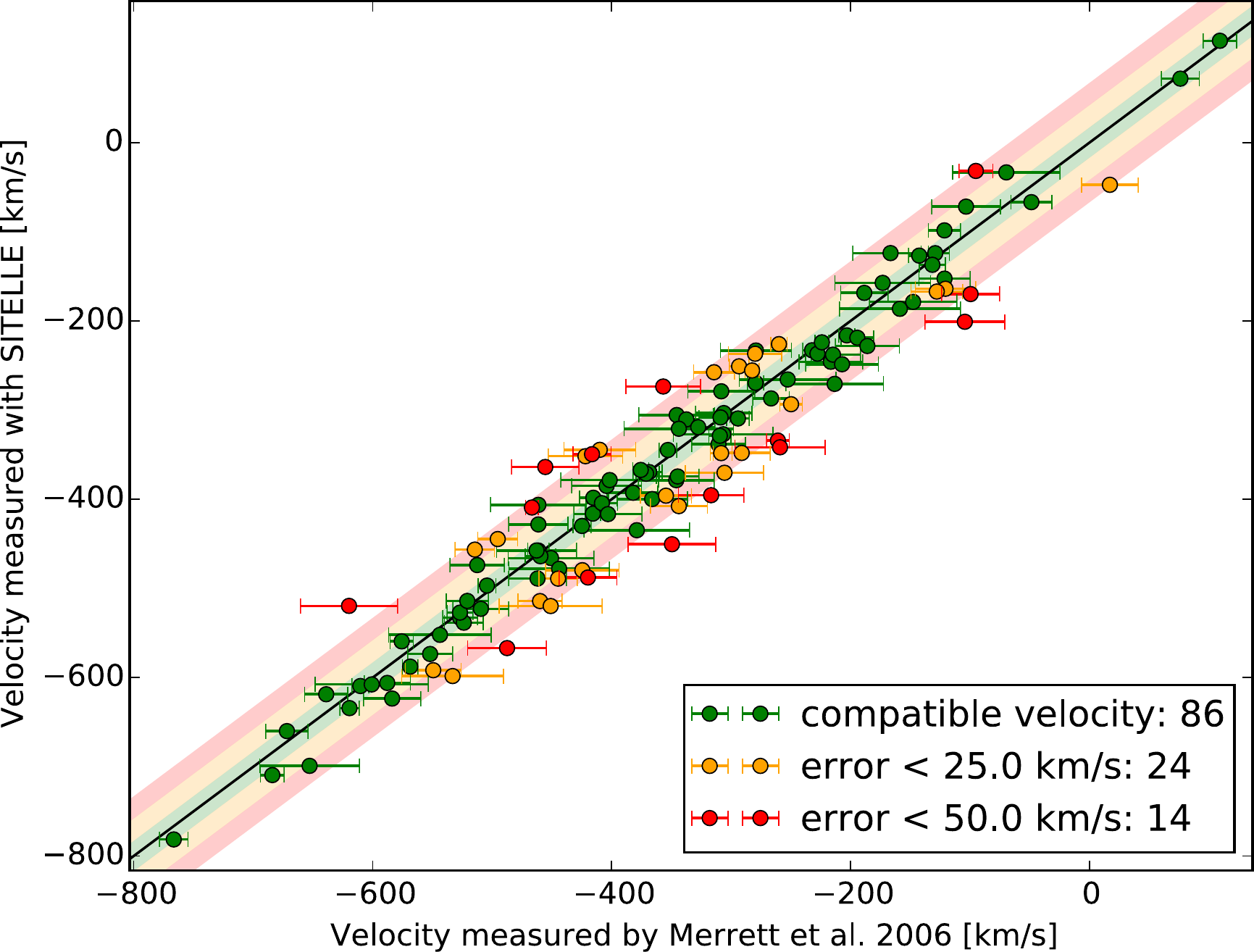}
 \caption{Comparison of the measured velocity of 124 planetary nebulae
   detected with SITELLE in M\,31 with the measurement of
   \citet{Merrett2006}. The resolution of the cube is 400. The
   one-to-one line is indicated by a black line.}
  \label{martin1:m31-pn-vel}
\end{figure}

The quality of the calibration has been checked by comparing the
velocity of 124 planetary nebul{\ae} (PNe) detected with SITELLE in
M\,31, from a low resolution data cube obtained during the
commissioning, with the velocity measured by \citet{Merrett2006}. 86
of the 124 PNe show a compatible velocity within the uncertainties
(see Figure~\ref{martin1:m31-pn-vel}).  A much more precise checking
has been obtained with an intermediate resolution spectral cube of the
same region obtained in August 2016 (see Martin et al. in
preparation). We have also compared the velocity map of M\,57 obtained
with SITELLE with the data obtained by \citet{Odell2007,Odell2013}
with an Echelle spectrograph \citep{Martin2016} and found to be in
good agreement with an overall precision below 0.5\,\kms on the
absolute and relative wavelength calibration.

Another source of absolute calibration uncertainty is the lack of
precision on the calibration laser wavelength. The error on the
velocity measurement $\epsilon_v$ is related to the error on the calibration laser wavelength, $\epsilon_\sigma$, since
\begin{equation}
\epsilon_v = c \frac{\epsilon_\sigma}{\sigma_\text{laser}}\;,
\end{equation}
with $\sigma_\text{laser}$ the real wavelength of the calibration
laser. Therefore, an error of 1 \AA{} on the calibration laser
wavelength translates into an error of 55\,\kms{}. This bias is easy
to correct since the measurement of the Meinel OH bands in a few cubes
is enough to obtain a better measurement. For the data release 1 we
have used the manufacturer value of 543.5\,nm which appears to be
biased by 80$\pm$5\,\kms{}.

\section{Astrometric calibration}
\label{sec:astrometric_calibration}

\begin{figure}
 \centering
 \includegraphics[width=\linewidth]{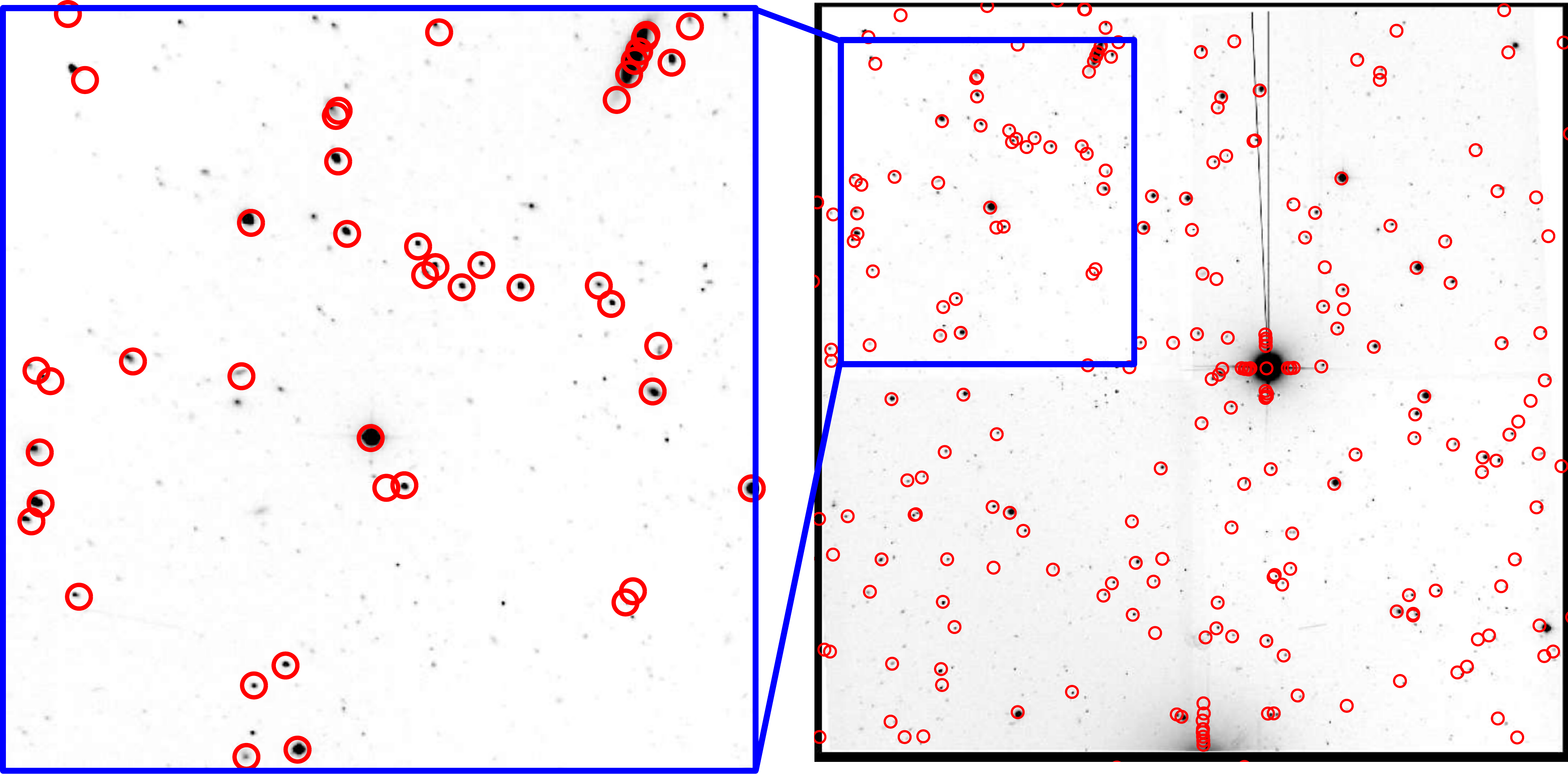}
 \caption{Positions of the stars from the USNO-B1 catalog transformed
   with the computed World Coordinate System (WCS) of the field around
   the planetary nebula M1-71.}
  \label{martin1:wcs}
\end{figure}

Astrometric calibration is computed from the fit of the point-like
sources detected in the field-of-view and the transformation of their
celestial coordinates \citep{Greisen2002} found in the USNO-B1 catalog
\citep{Monet2003}. The quality and the number of sources of the more
recent Gaia data release~1 catalog has motivated its use for the next
release instead of the old USNO catalog \citep{Gaia2016}. The fitting
engine fits all the stars at the same time which enhances the
precision of the transformation parameters. The astrometric
calibration is limited to 3\,pixels ($\sim$1'') in an 11 arc-minutes
circle around the center of the field by the optical distortions which
are not taken into account in the present data release (see
Figure~\ref{martin1:wcs}).

\section{Conclusions}
 
We have discussed the calibration quality of SITELLE's first data
release. We have shown that the absolute flux calibration was biased
by -5\% and that it was subject to a 5\% variability from one
observation to another. The general bias is likely to be corrected in
the next release via a more precise evaluation of the modulation
efficiency. But this results must be tempered with the fact that major
flux calibration biases cannot be avoided. We thus recommend that the
flux calibration be checked against external data. A $\sim$2\,\%
pixel-to-pixel error is expected on the basis of a comparison with an
Hubble images of M\,57. The absolute wavelength calibration is also
biased by 80$\pm$5\,\kms{} due to the lack of precision on the
calibration laser wavelength. The pixel-to-pixel error on the
calibration can be as large as 15\,\kms{} but it can be easily
corrected by measuring the velocity of Meinel OH bands in the
cube. This operation can be done with ORCS \citep{Martin2015}. The
astrometric calibration is done via the comparison with the USNO-B1
catalog and is limited to $\sim$1'' by the optical distortions which
are not corrected in the present release. All the observed biases will
be corrected in the next release. The precision on the pixel-to-pixel
wavelength calibration will also be enhanced by the analysis of the
internal phase of each cube that is directly related to the angle of
the incident light and therefore to the velocity calibration (Martin
et al., in preparation). The precision of the pixel-to-pixel flux
calibration will also be enhanced by using a 3D phase correction and a
better flatfield correction.

\section*{Acknowledgements}

This paper is based on observations obtained with SITELLE, a joint
project of Universit{\'e} Laval, ABB, Universit{\'e} de Montr{\'e}al
and the Canada-France-Hawaii Telescope (CFHT) which is operated by the
National Research Council (NRC) of Canada, the Institut National des
Science de l'Univers of the Centre National de la Recherche
Scientifique (CNRS) of France, and the University of Hawaii. LD is
grateful to the Natural Sciences and Engineering Research Council of
Canada, the Fonds de Recherche du Qu{\'e}bec, and the Canadian
Foundation for Innovation for funding.





\bibliographystyle{mnras}
\bibliography{calib-dr1} 

\begin{thebibliography}{}
\makeatletter
\relax
\def\mn@urlcharsother{\let\do\@makeother \do\$\do\&\do\#\do\^\do\_\do\%\do\~}
\def\mn@doi{\begingroup\mn@urlcharsother \@ifnextchar [ {\mn@doi@}
  {\mn@doi@[]}}
\def\mn@doi@[#1]#2{\def\@tempa{#1}\ifx\@tempa\@empty \href
  {http://dx.doi.org/#2} {doi:#2}\else \href {http://dx.doi.org/#2} {#1}\fi
  \endgroup}
\def\mn@eprint#1#2{\mn@eprint@#1:#2::\@nil}
\def\mn@eprint@arXiv#1{\href {http://arxiv.org/abs/#1} {{\tt arXiv:#1}}}
\def\mn@eprint@dblp#1{\href {http://dblp.uni-trier.de/rec/bibtex/#1.xml}
  {dblp:#1}}
\def\mn@eprint@#1:#2:#3:#4\@nil{\def\@tempa {#1}\def\@tempb {#2}\def\@tempc
  {#3}\ifx \@tempc \@empty \let \@tempc \@tempb \let \@tempb \@tempa \fi \ifx
  \@tempb \@empty \def\@tempb {arXiv}\fi \@ifundefined
  {mn@eprint@\@tempb}{\@tempb:\@tempc}{\expandafter \expandafter \csname
  mn@eprint@\@tempb\endcsname \expandafter{\@tempc}}}

\bibitem[\protect\citeauthoryear{Baril et~al.,}{Baril et~al.}{2016}]{Baril2016}
Baril M.~R.,  et~al., 2016, in Evans C.~J.,  Simard L.,   Takami H.,  eds,
  Proceedings of SPIE. International Society for Optics and Photonics, p.
  990829, \mn@doi{10.1117/12.2232075}

\bibitem[\protect\citeauthoryear{Bell}{Bell}{1972}]{Bell1972}
Bell R.~J.,  1972, {Introductory Fourier Transform Spectroscopy}.
Academic Press, New York, New York, USA

\bibitem[\protect\citeauthoryear{Bernier}{Bernier}{2006}]{Bernier2006}
Bernier A.-P.,  2006, in Proceedings of SPIE. SPIE, pp 626949--626949--9,
  \mn@doi{10.1117/12.671410}

\bibitem[\protect\citeauthoryear{Buton et~al.,}{Buton et~al.}{2012}]{Buton2012}
Buton C.,  et~al., 2012, \mn@doi [Astronomy {\&} Astrophysics]
  {10.1051/0004-6361/201219834}, 549, A8

\bibitem[\protect\citeauthoryear{Davis, Abrams  \& Brault}{Davis
  et~al.}{2001}]{Davis2001}
Davis S.~P.,  Abrams M.~C.,   Brault J. W. J.~W.,  2001, {Fourier transform
  spectrometry}.
Academic Press, San Diego

\bibitem[\protect\citeauthoryear{Doug}{Doug}{1993}]{TodyDoug1993}
Doug T.,  1993, in Hanisch R.~J.,  Brissenden R. J.~V.,   Barnes J.,  eds, ~
  Vol. 52, Astronomical Data Analysis Software and Systems II. Astronomical
  Society of the Pacific, p.~173

\bibitem[\protect\citeauthoryear{Drissen, Bernier, Rousseau-Nepton, Alarie,
  Robert, Joncas, Thibault  \& Grandmont}{Drissen et~al.}{2010}]{Drissen2010}
Drissen L.,  Bernier A.-P.,  Rousseau-Nepton L.,  Alarie A.,  Robert C.,
  Joncas G.,  Thibault S.,   Grandmont F.,  2010, in McLean I.~S.,  Ramsay
  S.~K.,   Takami H.,  eds, ~ Vol. 7735, SPIE Astronomical Telescopes +
  Instrumentation. International Society for Optics and Photonics, pp
  77350B--77350B--10, \mn@doi{10.1117/12.856470}

\bibitem[\protect\citeauthoryear{{Gaia Collaboration}}{{Gaia
  Collaboration}}{2016}]{Gaia2016}
{Gaia Collaboration} G.,  2016, \mn@doi [eprint arXiv:1609.04153]
  {10.1051/0004-6361/201629272}

\bibitem[\protect\citeauthoryear{Grandmont}{Grandmont}{2003}]{Grandmont2003}
Grandmont F.,  2003, in Proceedings of SPIE. SPIE, pp 392--401,
  \mn@doi{10.1117/12.457339}

\bibitem[\protect\citeauthoryear{Grandmont}{Grandmont}{2006}]{Grandmont2006}
Grandmont F.,  2006, PhD thesis, Universit{\'{e}} Laval

\bibitem[\protect\citeauthoryear{Grandmont, Drissen, Mandar, Thibault  \&
  Baril}{Grandmont et~al.}{2012}]{Grandmont2012}
Grandmont F.,  Drissen L.,  Mandar J.,  Thibault S.,   Baril M.~R.,  2012, in
  McLean I.~S.,  Ramsay S.~K.,   Takami H.,  eds, ~ Vol. 8446, SPIE -
  Ground-based and Airborne Instrumentation for Astronomy IV. International
  Society for Optics and Photonics, p. 84460U, \mn@doi{10.1117/12.926782}

\bibitem[\protect\citeauthoryear{Greisen \& Calabretta}{Greisen \&
  Calabretta}{2002}]{Greisen2002}
Greisen E.~W.,  Calabretta M.~R.,  2002, \mn@doi [Astronomy and Astrophysics]
  {10.1051/0004-6361:20021326}, 395, 1061

\bibitem[\protect\citeauthoryear{Hill et~al.,}{Hill et~al.}{2008}]{Hill2008}
Hill G.~J.,  et~al., 2008, in Kodama T.,  Toru Y.,   Aoki K.,  eds,  ASP
  Conference Series Vol. 399, Panoramic Views of Galaxy Formation and
  Evolution. Astronomical Society of the Pacific, p.~115

\bibitem[\protect\citeauthoryear{Kelz et~al.,}{Kelz et~al.}{2006}]{Kelz2006}
Kelz A.,  et~al., 2006, \mn@doi [Publications of the Astronomical Society of
  the Pacific] {10.1086/497455}, 118, 129

\bibitem[\protect\citeauthoryear{Learner, Thorne, Wynne-Jones, Brault  \&
  Abrams}{Learner et~al.}{1995}]{Learner1995}
Learner R. C.~M.,  Thorne A.~P.,  Wynne-Jones I.,  Brault J.~W.,   Abrams
  M.~C.,  1995, \mn@doi [Journal of the Optical Society of America A]
  {10.1364/JOSAA.12.002165}, 12, 2165

\bibitem[\protect\citeauthoryear{Maillard, Drissen, Grandmont  \&
  Thibault}{Maillard et~al.}{2013}]{Maillard2013}
Maillard J.~P.,  Drissen L.,  Grandmont F.,   Thibault S.,  2013, \mn@doi
  [Experimental Astronomy] {10.1007/s10686-013-9330-9}, 35, 527

\bibitem[\protect\citeauthoryear{Martin}{Martin}{2015}]{Martin2015t}
Martin T.,  2015, Phd thesis, Universit\'{e} Laval

\bibitem[\protect\citeauthoryear{Martin \& Drissen}{Martin \&
  Drissen}{2016}]{martin2016b}
Martin T.,  Drissen L.,  2016, in Reyl{\'{e}} C.,  Richard J.,  Cambr{\'{e}}sy
  L.,  Deleuil M.,  P{\'{e}}contal E.,  Tresse L.,   Vauglin I.,  eds,
  Proceedings of the annual meeting of the French Society of Astronomy {\&}
  Astrophysics Lyon, June 14-17, 2016June 14-17, 2016. No. October in
  Proceedings of the annual meeting of the French Society of Astronomy {\&}
  Astrophysics.
pp 23--28

\bibitem[\protect\citeauthoryear{Martin, Drissen  \& Joncas}{Martin
  et~al.}{2012}]{Martin2012}
Martin T.,  Drissen L.,   Joncas G.,  2012, in Radziwill N.~M.,  Chiozzi G.,
  eds, ~ Vol. 2, SPIE - Software and Cyberinfrastructure for Astronomy II. pp
  84513K--84513K--9, \mn@doi{10.1117/12.925420}

\bibitem[\protect\citeauthoryear{Martin, Drissen  \& Joncas}{Martin
  et~al.}{2015}]{Martin2015}
Martin T.,  Drissen L.,   Joncas G.,  2015, Astronomical Data Analysis Software
  an Systems XXIV (ADASS XXIV), 495

\bibitem[\protect\citeauthoryear{Martin, Prunet  \& Drissen}{Martin
  et~al.}{2016}]{Martin2016}
Martin T.~B.,  Prunet S.,   Drissen L.,  2016, eprint arXiv:1608.05854

\bibitem[\protect\citeauthoryear{Merrett et~al.,}{Merrett
  et~al.}{2006}]{Merrett2006}
Merrett H.~R.,  et~al., 2006, \mn@doi [Monthly Notices of the Royal
  Astronomical Society, Volume 369, Issue 1, pp. 120-142.]
  {10.1111/j.1365-2966.2006.10268.x}, 369, 120

\bibitem[\protect\citeauthoryear{Monet et~al.,}{Monet et~al.}{2003}]{Monet2003}
Monet D.~G.,  et~al., 2003, \mn@doi [The Astronomical Journal]
  {10.1086/345888}, 125, 984

\bibitem[\protect\citeauthoryear{O'Dell, Sabbadin  \& Henney}{O'Dell
  et~al.}{2007}]{Odell2007}
O'Dell C.~R.,  Sabbadin F.,   Henney W.~J.,  2007, \mn@doi [The Astronomical
  Journal] {10.1086/521823}, 134, 1679

\bibitem[\protect\citeauthoryear{O'Dell, Ferland, Henney  \& Peimbert}{O'Dell
  et~al.}{2013}]{Odell2013}
O'Dell C.~R.,  Ferland G.~J.,  Henney W.~J.,   Peimbert M.,  2013, \mn@doi [The
  Astronomical Journal] {10.1088/0004-6256/145/4/92}, 145, 92

\bibitem[\protect\citeauthoryear{Rousseau-Nepton}{Rousseau-Nepton}{2017}]{RNepton2017t}
Rousseau-Nepton L.,  2017, Phd thesis, Universit\'{e} Laval

\bibitem[\protect\citeauthoryear{Sakai, Vanasse  \& Forman}{Sakai
  et~al.}{1968}]{Sakai1968}
Sakai H.,  Vanasse G.~A.,   Forman M.~L.,  1968, \mn@doi [Journal of the
  Optical Society of America] {10.1364/JOSA.58.000084}, 58, 84

\bibitem[\protect\citeauthoryear{Sanchez, Rosales-Ortega, Kennicutt, Johnson,
  Diaz, Pasquali  \& Hao}{Sanchez et~al.}{2010}]{Sanchez2010}
Sanchez S.~F.,  Rosales-Ortega F.~F.,  Kennicutt R.~C.,  Johnson B.~D.,  Diaz
  A.~I.,  Pasquali A.,   Hao C.~N.,  2010, \mn@doi [Monthly Notices of the
  Royal Astronomical Society, Volume 410, Issue 1, pp. 313-340.]
  {10.1111/j.1365-2966.2010.17444.x}, 410, 313

\bibitem[\protect\citeauthoryear{Shara, Drissen, Martin, Alarie  \&
  Stephenson}{Shara et~al.}{2016}]{Shara2016}
Shara M.~M.,  Drissen L.,  Martin T.,  Alarie A.,   Stephenson F.~R.,  2016,
  \mn@doi [Monthly Notices of the Royal Astronomical Society]
  {10.1093/mnras/stw2753}, 465, 739

\bibitem[\protect\citeauthoryear{Wright, Corradi  \& Perinotto}{Wright
  et~al.}{2005}]{Wright2005}
Wright S.~a.,  Corradi R. L.~M.,   Perinotto M.,  2005, \mn@doi [Astronomy and
  Astrophysics] {10.1051/0004-6361:20052666}, 436, 9

\makeatother
\end{thebibliography}






\bsp	
\label{lastpage}
\end{document}